\documentclass[11pt]{article}

\usepackage{acl}

\usepackage{times}
\usepackage{latexsym}

\usepackage[T1]{fontenc}
\usepackage[utf8]{inputenc}
\usepackage{microtype}
\usepackage{inconsolata}
\usepackage{graphicx}
\usepackage{multirow}
\usepackage{xcolor}
\usepackage{colortbl}
\usepackage{enumitem}
\usepackage{arydshln}
\usepackage{amsmath}
\usepackage{algorithm}
\usepackage{algpseudocode}
\usepackage{quoting}
\usepackage{amssymb}
\usepackage{tcolorbox}
\usepackage[table]{xcolor}
\usepackage{array}
\usepackage{xspace}

\usepackage[T1]{fontenc}

\usepackage[utf8]{inputenc}

\usepackage{microtype}

\usepackage{inconsolata}

\usepackage{graphicx}

\newcommand{\todo}[1]{\textcolor{red}{#1}}
\newcommand{\new}[1]{\textcolor{black}{#1}}

\newcommand{\proposedmethod}{RAAC} 
\newcommand{\proposedcomponent}{controller} 

\newcommand{\miniskip}{\vspace*{-.5\baselineskip}}
\newcommand{\shrink}{\vspace*{-.9\baselineskip}}

\makeatletter
\DeclareRobustCommand\onedot{\futurelet\@let@token\@onedot}
\def\@onedot{\ifx\@let@token.\else.\null\fi\xspace}

\def\eg{e.g\onedot,\xspace} 

\makeatother

\title{\todo{On the Tendency of Deep Research Agents Toward Reasoning Stagnation}}
\title{When Deep Research Agents Get Stuck: Analyzing and controlling deep re}
\title{On Repetitive and Stagnant Reasoning in Deep Research Agents}
\title{Understanding and Controlling Stagnant Reasoning in  \\Deep Research Agents}

\title{Reasoning Stagnation in Deep Research – Overcoming with Agent-Agnostic Retrieval Sufficiency}

\title{Reasoning Stagnation in Deep Research Agents: Breaking the Loop with Retrieval Sufficiency}
\title{When Deep Research Reasoning Get Stuck: Augmenting Agents with Retrieval Sufficiency}

\title{Stagnant and Repetitive Reasoning in Deep Research: Agent-Agnostic Augmentation with Retrieval Sufficiency}

\title{Stagnation in Deep Research Agents: Enhanced Reasoning with Retrieval Sufficiency}

\title{Deep Research Agents Stagnate in Repetitions –\\ Enhanced Reasoning with Retrieval Sufficiency}

\title{Deep Research Agents Stagnate in Search!\\ Enhanced Reasoning with Retrieval Sufficiency}

\title{Stagnant Reasoning in Deep Research Agents Approached with Retrieval-Aware Controller}

\title{RAAC: Retrieval-Aware Controller to Break Stagnant Reasoning of\\ Deep Research Agents}
\title{Addressing Stagnant Research in Deep Research Agents with\\ Retrieval-Aware Controller}

\title{When Deep Research Agents Stagnate: Enhancing Reasoning with Retrieval-Aware Agent Control}

\author{
  \textbf{Heydar Soudani\textsuperscript{1,2}},
  \textbf{Elizabeth Lingg\textsuperscript{2}},
  \textbf{Faegheh Hasibi\textsuperscript{1}},
  \textbf{Navid Rekabsaz\textsuperscript{2}}
\\
  \textsuperscript{1}Radboud University\\
  \textsuperscript{2}Thomson Reuters Labs
\\
  \small{
    \textbf{Correspondence:} \href{mailto:}{heydar.soudani@ru.nl}
  }
\\
\texttt{\{heydar.soudani,faegheh.hasibi\}@ru.nl} 
\\
\texttt{\{elizabeth.lingg,navid.rekabsaz\}@thomsonreuters.com} 
}

\begin{document}
\maketitle

\begin{abstract}
In this paper, we analyze the reasoning trajectories of a variety of DRAs and show that existing agents often suffer from \emph{reasoning stagnation}: the majority of iterations contribute little or no improvement to final performance, while agents lack awareness of their trajectories and are therefore ineffective at adapting their search strategies or determining when to terminate. To address this issue, we introduce a set of unsupervised signals and a Retrieval-Aware Agent Controller (RAAC), which assists the agent in selecting optimal actions at each stage of the research process.
 RAAC incorporates key information retrieval principles, namely \emph{search novelty} and \emph{information coverage}, resulting in more effective reasoning trajectories that improve overall performance while reducing unnecessary iterations, and consequently cost and latency.
Specifically on \textsc{BrowseComp-Plus} and across a large set of DRAs, adding RAAC reduces the number of search calls by an average of 14, significantly improves the best-performing DRA on recall and accuracy, and achieves an accuracy gain of up to 10\% (3\% on average).

\end{abstract}

\section{Introduction}
\label{sec:introduction}

\begin{figure}[h]
  \centering
  \includegraphics[width=0.46\textwidth]{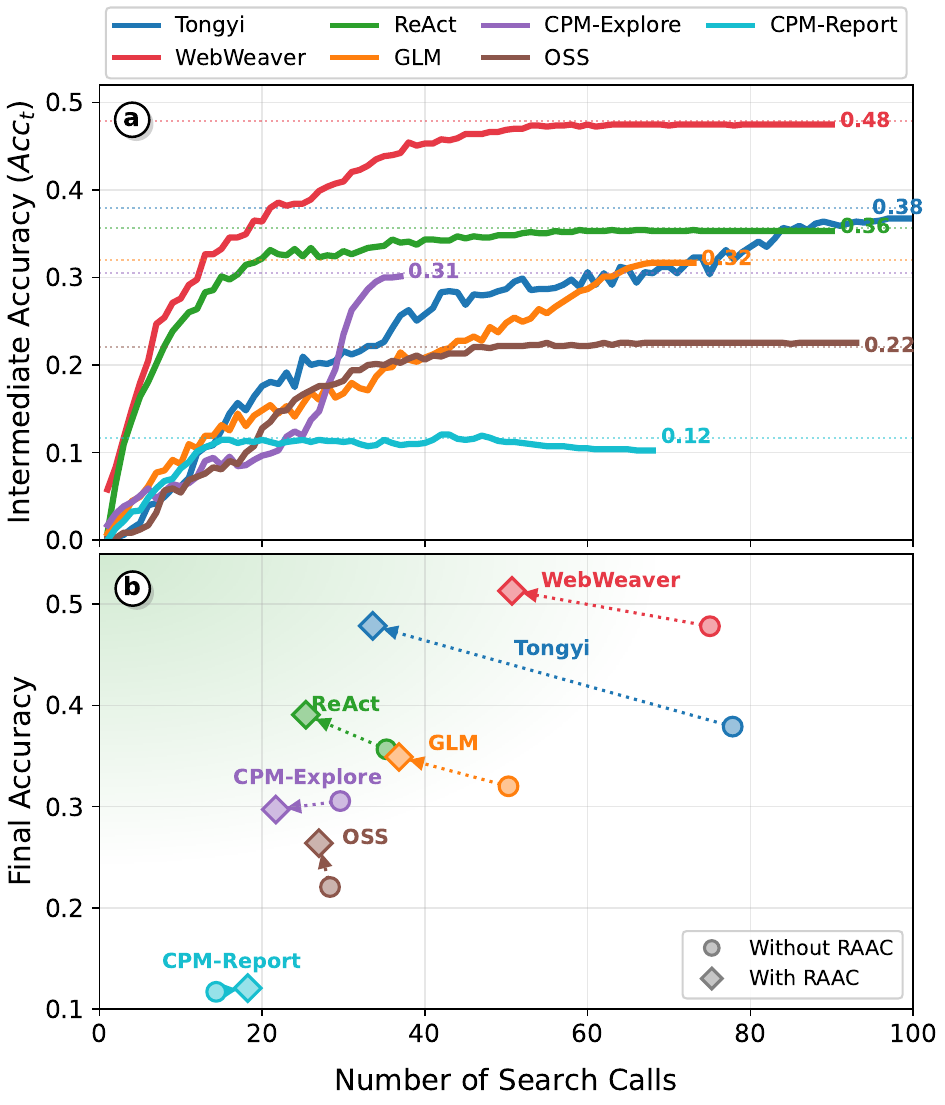}
  \miniskip
  \caption{
  (a) \textbf{\textit{Stagnant reasoning}} in DRAs: the agents  continue for many unnecessary repetitive steps, even though their information discovery capabilities are saturated and the task often has already been accomplished.
  (b) Effect of adding \proposedmethod: the introduced unsupervised signals guide DRAs toward a more effective decision-making, improving both efficiency and task success. 
  }
  \label{fig:accumulative_accuracy}
  \shrink
\end{figure}


\begin{figure*}[t]
  \centering
  \includegraphics[width=0.94\textwidth]{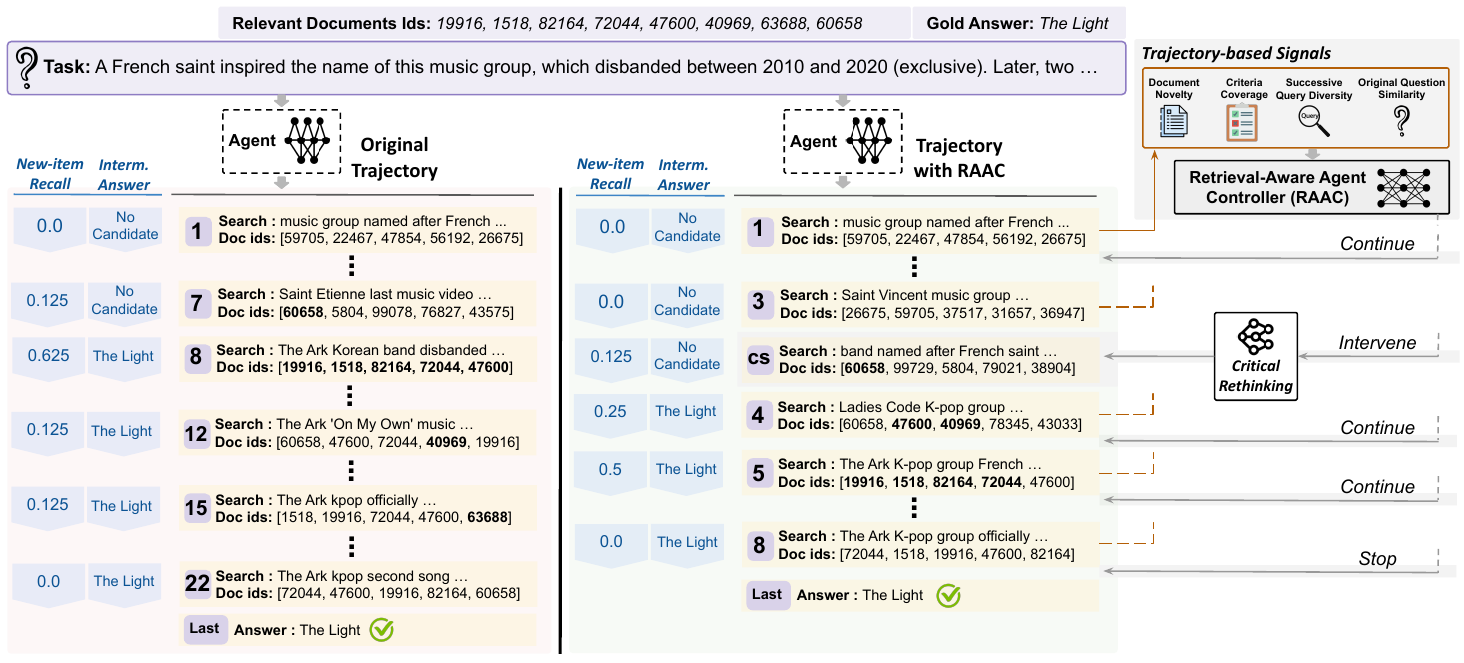}
  \miniskip
  \caption{
  Comparison of the trajectories of a given query between an arbitrary Deep Research Agent (DRA) with and without Retrieval-Aware Agent Controller (\proposedmethod). In the original trajectory, the agent answers the query in 23 iterations, although the answer becomes available at iteration 8. With \proposedmethod, the agent finds a relevant document earlier, making the answer available at iteration 4, and the controller stops the process at iteration 8.
  }
  \label{fig:example}
  \shrink
\end{figure*}

%
Retrieval-Augmented Generation (RAG) has become the dominant paradigm for grounding Large Language Model (LLM) outputs in external knowledge~\cite{Noise2024Florin, mallen23popqa, Soudani24FTvsRAG}.
Deep Research Agents (DRAs) extend RAG by combining tool use (\eg  search engines) with multi-step reasoning, implemented through either instruction-tuned agents~\cite{press2023selfask, Yao23ReAct, li25searcho1} or reinforcement learning-trained agents~\cite{Jin25SearchR1, Li25Tongyi}. 

Recent work on DRAs has largely focused on component-level improvements, such as enhancing retrieval through re-ranking and query reformulation techniques~\cite{chen2026agentir,meng2026revisiting,sharifymoghaddam2026rerank}, or strengthening reasoning by incorporating additional stages such as outline generation and report writing~\cite{li2026agentcpm,li2025webweaver}. However, a critical question remains unexplored: \textit{How effectively do these components work together over long reasoning trajectories to achieve the final objective?}

In this work, we shift the focus to the intermediate steps of the agent's trajectory and aim to uncover failure patterns in DRAs, irrespective of their underlying design choices. We systematically evaluate seven DRAs with diverse architectures at each stage of their trajectories using intermediate recall and accuracy. Our analysis reveals a consistent pattern: agents often lack awareness of the information they have already collected and whether they have reached a point where information discovery has saturated.  As a result, agents continue issuing search calls and generating reasoning steps while making little or no meaningful progress toward task completion; see Figure~\ref{fig:accumulative_accuracy}(a). We refer to this observed pattern as \textbf{stagnant reasoning}.

To address this issue, building on our analysis, we hypothesize that DRAs should maintain a broad view of their trajectories, particularly regarding what information has already been covered and whether subsequent steps explore new regions of the search space. This information is necessary for the agent to decide when to continue, redirect/reflect, or stop, preventing it from stagnating.
Based on this, we propose a set of unsupervised trajectory-based signals that capture coverage of query criteria and novelty of queries and documents.
We integrate these signals into DRAs by introducing \textbf{R}etrieval-\textbf{A}ware \textbf{A}gent \textbf{C}ontroller (\textbf{\proposedmethod}), an agent-agnostic decision-making controller. 
At each step of the research process, \proposedmethod\ leverages these signals to guide the agent to \textit{continue} its reasoning trajectory, \textit{intervene} and adjust its course, or \textit{stop} and generate the final answer.
\proposedmethod\ is agent-agnostic and independent of the design/complexity of the underlying DRA (cf. Figure~\ref{fig:example} 
for an example.)

Our experiments show that applying \proposedmethod\ to DRAs yields more optimized reasoning trajectories; see Fig.~\ref{fig:accumulative_accuracy}(b). Agents with \proposedmethod\ take fewer steps and execute fewer search calls than those without, lowering computational cost and latency. In particular, \proposedmethod\ reduces search calls by 14 points on \textsc{BrowseComp-Plus}~\cite{chen2025browsecompplus} (up to 44), 4 points on \textsc{NeuCLIR}~\cite{lawrie2025neuclir}, and 5.5 points on \textsc{LegalSearch}~\cite{korikov2025batched}. 
Despite this, the agents maintain or even improve their performance, achieving accuracy gains of up to 10\% and an average improvement of 3\% across all evaluated agents on \textsc{BrowseComp-Plus}.
We further analyze the effect of our proposed controller on DRAs through correlation and ablation studies, highlighting the importance of retrieval-aware signals in DRAs.

\noindent
In summary, our contributions are as follows:
\begin{itemize}[leftmargin=*, itemsep=2pt, topsep=2pt]
   \item We conduct, to our knowledge, the first systematic analysis of DRAs based on intermediate behavior, uncovering the \textit{\textbf{reasoning stagnation}} problem, where the model lacks awareness of its progress and cannot adapt its search direction or decide when to stop.
   \item We propose unsupervised trajectory-based signals to mitigate stagnant reasoning and  steer DRAs toward more effective and efficient reasoning trajectories.
   \item We demonstrate the effectiveness of these signals and design a controller that regulates the reasoning process, improving both the efficiency and effectiveness of DRA agents and helping guide future research in DRA optimization.
\end{itemize}

\section{Related Work}~\label{sec:related_work}
\noindent
\textbf{Deep Research Agents: }
Retrieval-augmented generation has evolved from a simple retrieve-then-generate paradigm~\cite{Noise2024Florin, Soudani24FTvsRAG} to approaches that more tightly integrate retrieval into generation, often interleaving it with intermediate reasoning steps~\cite{trivedi2023ircot, jiang2023flare, Hoveyda2025, Yao23ReAct}. This shift, further driven by Reinforcement Learning (RL), has led to Deep Research Agents (DRAs), which perform multi-step tool use and extensive information seeking to handle complex queries~\cite{Jin25SearchR1, rafiee2026trqa}. 
Recent work on improving DRAs has focused on strengthening the search component via retrieval fine-tuning~\cite{chen2026agentir}, reranking~\cite{sharifymoghaddam2026rerank}, and query reformulation~\cite{meng2026revisiting}, often leveraging signals from reasoning traces.




A key distinguishing feature of DRAs is their agentic workflow, i.e., how tools are selected, ordered, and used. We group DRAs into four categories based on how this workflow is induced. \textbf{The first and second groups}, instruction-tuned~\cite{press2023selfask, li25searcho1} and RL-trained methods~\cite{openai2025gptoss, glm2025arc, Li25Tongyi, Chen26cpmexplore, Shao25drtulu}, both follow a ReAct-style loop~\cite{Yao23ReAct}, which can be described as:
\begin{tcolorbox}[colback=gray!10, colframe=gray!10, boxrule=0pt, arc=1mm, boxsep=0pt, before skip=4pt, after skip=4pt]
{\fontsize{9pt}{14pt}\selectfont
Query $\rightarrow$ [Think $\rightarrow$ Search $\rightarrow$ Observe]* $\rightarrow$ Answer,
}
\end{tcolorbox}

\noindent
where $*$ denotes that the loop is repeated for as many iterations as the agent deems necessary.
\textbf{The third group} follows an outline-guided search paradigm~\cite{han25DeepResearcher}. WebWeaver~\cite{li2025webweaver} introduces a dynamic outline-guided workflow consisting of a planner and a writer. In the planning phase, the model generates an outline along with a memory bank. In the writing phase, the writer retrieves information from the memory bank to produce the final report, as follows:
\begin{tcolorbox}[colback=gray!10, colframe=gray!10, boxrule=0pt, arc=1mm, boxsep=0pt, before skip=4pt, after skip=4pt]
\begin{minipage}{\linewidth}
{\fontsize{9pt}{12pt}\selectfont
Query $\rightarrow$ 

[Think $\rightarrow$ Search $\rightarrow$ Write\_outline]* $\rightarrow$ Outline $\rightarrow$

[Think $\rightarrow$ Retrieve $\rightarrow$ Write\_section]* $\rightarrow$ Report
}
\end{minipage}
\end{tcolorbox}

\noindent
\textbf{The fourth group} adopts a report-based generation style. AgentCPM-Report~\cite{li2026agentcpm} introduces a Writing-as-Reasoning paradigm, in which the report is generated incrementally throughout the search process rather than being preceded by a standalone outline. In addition, the outline (or plan) can be dynamically revised during generation when necessary. The overall workflow is as follows:
\begin{tcolorbox}[colback=gray!10, colframe=gray!10, boxrule=0pt, arc=1mm, boxsep=0pt, before skip=4pt, after skip=4pt]
\begin{minipage}{\linewidth}
{\fontsize{9pt}{12pt}\selectfont
Query $\rightarrow$ Search $\rightarrow$ Init Plan $\rightarrow$ [Search $\rightarrow$ Write]* $\rightarrow$ 

[Extend Plan $\rightarrow$ [Search $\rightarrow$ Write]*]*  $\rightarrow$ Report
}
\end{minipage}
\end{tcolorbox}

\noindent
\new{Moreover, parallel-based approaches have also been proposed, such as FlashSearcher~\cite{Qin26FlashSearcher} and ParallelResearch~\cite{nie2025efficient}, which first decompose a task into multiple subtasks and then execute a DRA pipeline for each subtask. Nevertheless, the workflow associated with each subtask can still belong to one of the four groups.}

In this paper, we investigate DRAs across all four groups to demonstrate the broad applicability of our findings.
\new{Specifically, instead of proposing a new agent workflow, we focus on leveraging trajectory-aware signals through a lightweight controller to mitigate the stagnant reasoning behavior of DRAs.}

\medskip
\noindent
\textbf{Retrieval Control:}
Signal-driven controllers have recently been proposed for adaptive RAG systems. FLARE~\cite{jiang2023flare} and DRAGIN~\cite{su2024dragin} trigger retrieval based on token-level confidence, Adaptive-RAG~\cite{jeong2024adaptiverag} uses a complexity classifier for query routing, and Stop-RAG~\cite{park2025stop} learns when to stop retrieval through a finite-horizon MDP. \new{DeepControl~\cite{Xiong2026deepcontrol} and InfoGain-RAG~\cite{Wang25InfoGainRAG} estimate the utility of retrieved information to guide adaptive retrieval and document filtering.} However, these methods typically rely on a single control signal and support only binary decisions. 
\new{In contrast, our controller integrates multiple trajectory-level signals that provide interpretable views of different trajectory states, rather than relying on a single learned utility score, and supports a richer action space.}

\section{Stagnant Reasoning Analysis}\label{sec:dra_analysis}
%
We systematically analyze DRAs across iterations to understand their stagnant reasoning behavior. In what follows, we first provide a formal definition of DRAs, then explain the method used for analyzing intermediate search and reasoning behavior of DRAs, and finally report the results of the analysis.

\setlength{\abovedisplayskip}{2pt} 
\setlength{\belowdisplayskip}{2pt} 

\begin{figure*}[t]
  \centering
  \shrink
  \includegraphics[width=0.98\textwidth]{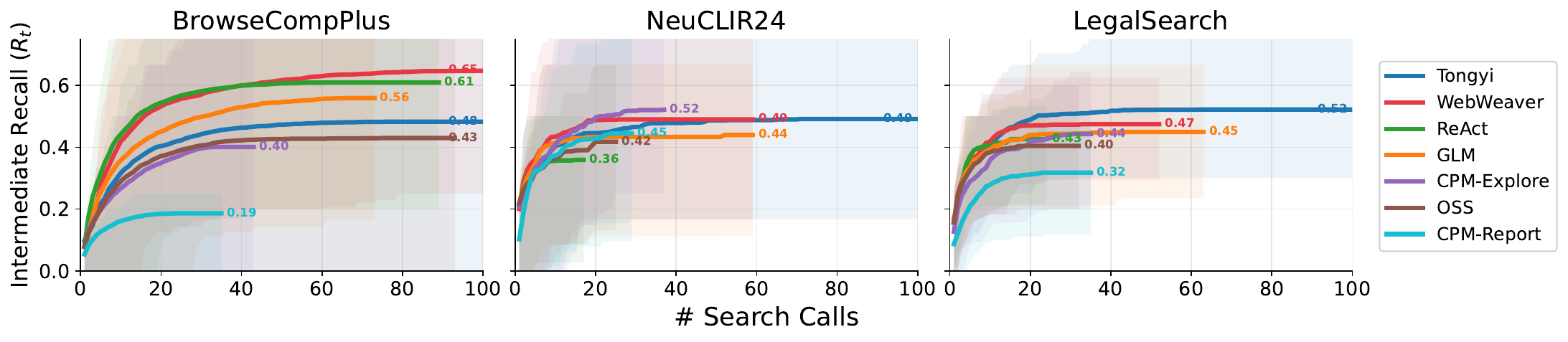}
  \miniskip
  \caption{Intermediate Recall ($R_t$) across seven agents and three datasets, illustrating the stagnation problem in DRAs. Shaded bands indicate the spread (standard deviation) across queries. 
  }
  \label{fig:accumulative_recall}
  \shrink
\end{figure*}

\subsection{Deep Research – Task Formulation}

Following prior work~\cite{Li25Tongyi, li2025webweaver, chen2026agentir}, we formulate DRAs based on the ReAct framework~\cite{Yao23ReAct}, which combines reasoning and acting. 
In this framework, an LLM agent interacts with the environment (e.g., a retriever) over multiple turns. At each turn $t$,  the agent generates both a thought $\tau_t$, representing the agent's internal reasoning process, and a subsequent action $a_t$, representing an external operation performed to interact with the environment. 
After executing the action $a_t$, the agent receives the feedback $o_t$ from the environment.
This process forms a trajectory, denoted as $\mathcal{H}_T$, consisting of a sequence of thought--action--observation triplets:
$$ \mathcal{H}_T = (\tau_1, a_1, o_1, \cdots, \tau_t, a_t, o_t, \cdots, \tau_T, a_T).$$
Each thought $\tau_t$ and action $a_t$ are generated by the agent's policy $\pi$, conditioned on a representation of the interaction history $\mathcal{H}_{t-1}$:
$$\tau_t, a_t \sim \pi(\cdot \mid f(\mathcal{H}_{t-1})),$$
where the function $f$ generates a representation of the interaction history, for example by concatenating all text in $\mathcal{H}_{t-1}$~\cite{Yao23ReAct, tongyi2025deepresearch} or by producing a concise summary of the agent's previous thoughts and observations ~\cite{li2025webweaver, li2026agentcpm}.

In the most common setting, where the environment consists of a single retriever, each intermediate action corresponds to a search query $q_t$ (i.e., $a_t = q_t$), the final action corresponds to the generated response $r_T$ (i.e., $a_T = r_T$).
In this case, the observation $o_t$ consists of the top-$k$ retrieved documents $D_t$ (i.e., $o_t = D_t$), which are obtained by the retrieval model $\mathcal{R}$; i.e., $D_t \leftarrow \mathcal{R}(q_t)$.


\subsection{Our Analysis Approach}~\label{sec:approach}
To better understand the DRAs' decision-making processes, we evaluate their reasoning and search behavior across intermediate turns using a set of evaluation methods described below.

\medskip 
\noindent\textbf{Intermediate Accuracy.}
DRAs commonly generate the final answer $r_T$ only at the end of the trajectory, after gathering sufficient information. To assess whether a given agent can already answer the user query at iteration $t$, we spawn a conceptual branch from the main reasoning trajectory by prompting the agent to generate a secondary reasoning step $\tau_t^{\prime}$, which leads to an intermediate answer $r_t$ (i.e., $a_t^{\prime} = r_t$) without interfering with the main trajectory. We also allow the agent to output \textit{no candidate} if sufficient information has not yet been gathered. More specifically, two separate trajectories are generated at turn $t$:
$$\tau_t^{\prime}, r_t \sim \pi(\cdot \mid f(\mathcal{H}_{t-1})),$$ 
$$\tau_t, q_t \sim \pi(\cdot \mid f(\mathcal{H}_{t-1})),$$ 
where $r_t$ is either an answer to the user's question or \textit{no candidate} and $\tau_t^{\prime}$ is the thought leading to the answer $r_t$. The agent further continues its trajectory with the thought $\tau_t$ and query $q_t$.
%
We compute the intermediate accuracy (Acc$_t$) by comparing the intermediate response $r_t$ with the ground truth response $r^*$. 
See Appendix~\ref{sec:app:exp} for the further details of this metric.

\medskip 
\noindent\textbf{Intermediate Recall.}
To measure how effectively the retriever discovers relevant information throughout the reasoning trajectory, we define \textit{Intermediate Recall}. At step $t$, it is defined as the proportion of gold documents that have been retrieved up to that step:
$$\text{R}_t = \left| D_g \cap D_{1:t} \right| / |D_g|,$$
where $D_g$ denotes the set of relevant gold documents, and
$D_{1:t} = \bigcup_{i=1}^{t} D_i$ represents the union of the documents retrieved up to step $t$.

\medskip 
\noindent\textbf{Retrieved Document Types.}
To analyze the reasoning and search behavior of DRAs more precisely, we examine the retrieved documents and categorize them into four types.
Formally, each  document $d_t$ retrieved in step $t$ belongs to one of the following categories:

\medskip 
\noindent\textsc{Relevant New}: 
The document is labeled as a gold document and has not been seen before in the trajectory
    $d_t \in D_g$ and $ d_t \notin D_{1:t-1}.$

\medskip 
\noindent\textsc{Relevant Seen}: 
The document is a gold document and has already been observed in the trajectory; i.e.,
    $d_t \in D_g$ \text{and} $d_t \in D_{1:t-1}.$

\medskip 
\noindent\textsc{Irrelevant New}: The document is not a gold document and has not been seen before in the trajectory; i.e.,
    $d_t \notin D_g \quad \text{and} \quad d_t \notin D_{1:t-1}.$

\medskip 
\noindent\textsc{Irrelevant Seen}: The document is not a gold document but has already been observed in the trajectory, i.e.,
    $d_t \notin D_g \quad \text{and} \quad d_t \in D_{1:t-1}.$

\subsection{Experiments and Results}~\label{sec:analysis_results}
\noindent\textbf{Deep Research Agents.}
We evaluate seven DRAs, belonging to four different architectures (cf. Section~\ref{sec:related_work}): ReAct~\citep{Yao23ReAct}, GLM-4.7~\citep{glm2025arc}, GPT-oss-20b-high~\citep{openai2025gptoss}, Tongyi-DR~\citep{tongyi2025deepresearch}, AgentCPM-Explore~\citep{Chen26cpmexplore}, WebWeaver~\citep{li2025webweaver}, and AgentCPM-Report~\citep{li2026agentcpm}. For ReAct and WebWeaver, we use \verb|Claude Sonnet 4.5| as the backbone LLM.
To ensure fair comparison, all agents use the same retrieval pipeline and receive the top-$k$ retrieved documents per search ($k=5$), following~\citet{chen2026agentir, meng2026revisiting}; see Table~\ref{tab:agent_hyperparams} in Appendix~\ref{sec:app:exp} for further details.

\medskip 
\noindent\textbf{Datasets.}
We conduct experiments on three 
datasets: \textsc{BrowseComp-Plus}~\citep{chen2025browsecompplus}, \textsc{NeuCLIR}~\citep{lawrie2025neuclir}, and \textsc{LegalSearch}, a closed-source dataset provided by Thomson Reuters~\citep{korikov2025batched}. All datasets contain human-verified supporting documents, enabling the evaluation of retrieval performance. In addition, \textsc{BrowseComp-Plus} includes short verifiable answers, allowing us to evaluate both intermediate and final generated responses. See Appendix~\ref{sec:app:exp} for further details.

\medskip 
\noindent\textbf{Retriever.}
Following~\citet{chen2026agentir, meng2026revisiting}, we employ  a single-vector dense retriever using \verb|Qwen3-Embedding-4B|~\citep{qwen2025embedding} for \textsc{BrowseComp-Plus} and \textsc{NeuCLIR} datasets. 
For \textsc{LegalSearch}, documents are retrieved through a proprietary multi-stage retrieval pipeline.

\medskip 
\noindent\textbf{Evaluation Metrics.}
We report intermediate accuracy (Acc$_t$) and recall (R$_t$) on \textsc{BrowseComp-Plus} and intermediate recall on other datasets. 
For \emph{accuracy}, we follow the \textsc{BrowseComp-Plus} evaluation protocol ~\citep{chen2025browsecompplus} and leverage LLM-as-judge comparison of the agent's final answer against the ground truth.

\begin{figure*}[t]
  \centering
  \includegraphics[width=0.98\textwidth]{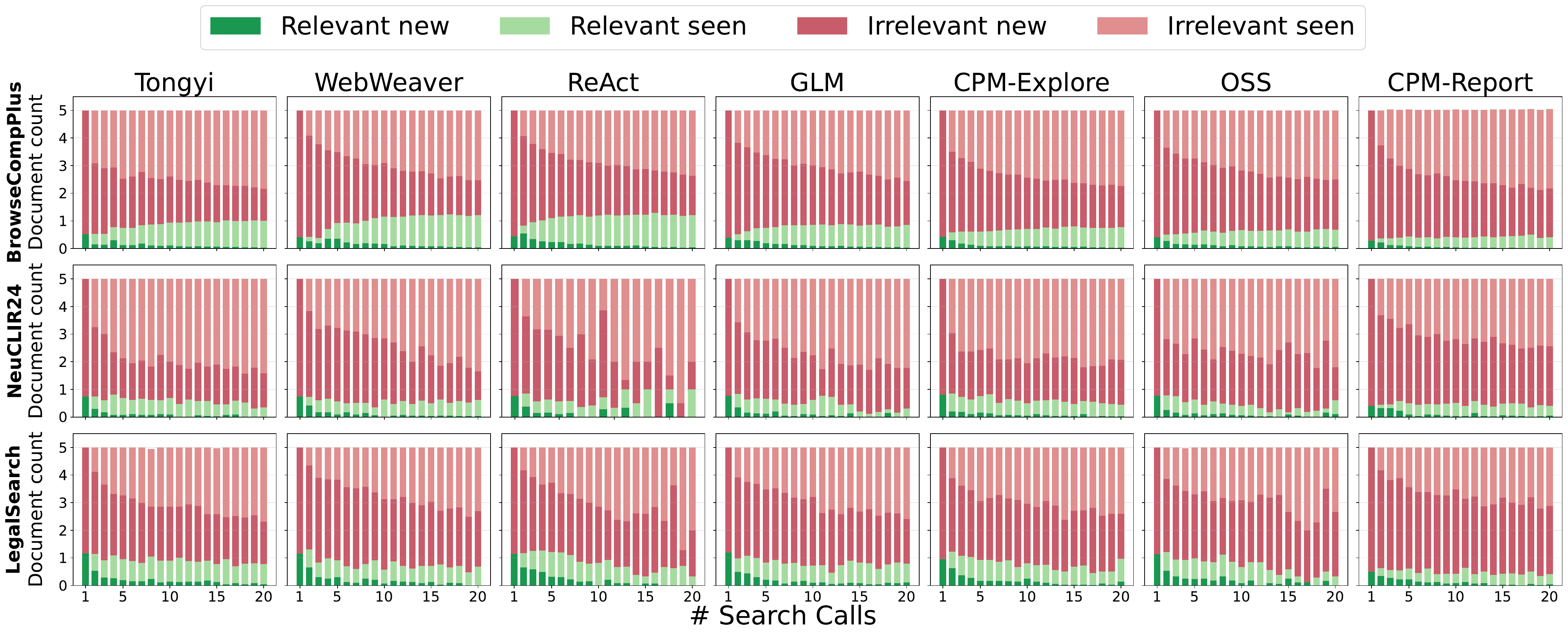}
  \miniskip
  \caption{Distribution of document types per reasoning step across the first 20 iterations, showing a reduction in newly retrieved relevant documents.}
  \label{fig:retrieved_docs_types_comparison}
  \shrink
\end{figure*}


\medskip
\noindent\textbf{(RQ1) 
How do DRAs' search and reasoning performances change throughout the reasoning trajectory?}
Figure~\ref{fig:accumulative_recall} presents the intermediate recall for all datasets. While the best performing DRAs varies across these datasets,  a consistent pattern emerges across all datasets and agents: after the initial steps, performance quickly saturates, and additional iterations provide little to no meaningful improvement in recall.
A similar behavior is observed for intermediate accuracy on \textsc{BrowseComp-Plus}. As shown in Figure~\ref{fig:accumulative_accuracy}, most agents (WebWeaver, ReAct, GPT-oss, and CPM-Report) achieve near-final performance after several iterations, with later steps adding little benefit. CPM-Explore also follows the trend of reaching its final accuracy within a few iterations but tends to stop once it finds the correct answer. Only GLM and Tongyi improve gradually, with each iteration contributing consistently to performance.


These results suggest that DRAs often suffer from reasoning stagnation, where agents continue generating steps without making meaningful progress toward task completion. During these reasoning steps, \textbf{agents are not fully aware of what they have already gathered}, which constraints have been satisfied, or whether they have already reached the correct final answer. In other words, they lack a clear sense of progress toward the solution. As a result, even after obtaining sufficient information or identifying the correct answer, they may continue invoking tools unnecessarily.


\medskip 
\noindent\textbf{(RQ2) How does the distribution of document types observed by the agent evolve across iterations?}
Figure~\ref{fig:retrieved_docs_types_comparison} shows the distribution of document types across different agents and datasets for the first 20 iterations of each DRA, which sufficiently captures the observed patterns. 
We observe a consistent pattern across all settings: over time, agents retrieve fewer new documents (both relevant and irrelevant) and increasingly revisit previously seen ones. Meanwhile, newly retrieved relevant documents steadily decline and eventually disappear after several iterations.

These patterns suggest that \textbf{agents struggle to retrieve alternative evidence needed for the final answer},
instead revisiting previously seen relevant documents. This suggests that reasoning stagnation stems from limited exploration of the search space and difficulty escaping local optima defined by the early relevant evidence.

\section{Retrieval-Aware Agent Controller}~\label{sec:signals_controller}
We attribute stagnant reasoning in DRAs to two main causes: (1) limited awareness of information coverage and (2) difficulty shifting criteria or exploring new aspects. We argue that agents need trajectory-level retrieval-aware signals to guide their reasoning process. We propose four unsupervised signals and show that integrating them into DRAs as a controller improves performance.

\subsection{Retrieval-Aware Signals}

\noindent
\textbf{Criteria Coverage (Cov).} 
Each deep research question is commonly associated with a set of criteria $C = [c_1, \ldots, c_M]$, stating what information is needed to fully accomplish the task~\cite{chen2025browsecompplus, min23FActScore}.
These criteria may be explicitly stated in the query (as for instance provided in \textsc{BrowseComp-Plus} dataset) or otherwise, should be inferred by the agent. 

We define Criteria Coverage as the extent to which these criteria have been satisfied over the course of research. The metric reflects whether the agent is effectively exploring the search space, becoming overly focused on a subset of criteria, or has gathered sufficient information to comprehensively answer the question. More specifically, the coverage score at iteration $t$ is defined as:
$$\mathrm{Cov}_{t} = \frac{1}{|C|} \sum_{c \in C} W(c, t),$$
where $W(c,t)$ is a weighting function representing the coverage of criterion $c$ at iteration $t$. Following~\citet{Support25Thakur, Enabling23Gao}, we assign weights of $0.5$, $1.0$, and $0$ to partially covered, fully covered, and uncovered criteria, respectively.

\medskip\noindent 
\textbf{Document Novelty (Nov).} This signal
measures the fraction of newly retrieved documents in an iteration that have not appeared in any previous iterations. This choice is motivated by the observation in the previous section, where the retrieval of new (relevant) documents declines in the further steps of the trajectory. 
Novelty at iteration $t$ is defined as the proportion of retrieved documents $D_t$ that do not appear in $D_{1:t-1}$, formulated below.
$$ \mathrm{Nov}_{t} = \frac{\left|D_t \setminus D_{1:t-1}\right|} {\left|D_t\right|}.$$ 

\medskip\noindent
\textbf{Successive Query Diversity (Div).}
As a complementary signal to \textit{document} novelty, we define a diversity signal for generated \textit{queries}. This signal measures the dissimilarity between the current search query and the query from the previous iteration, and is  defined as $\mathrm{Div}_{t} = 1 - \mathrm{cos}(e_t, e_{t-1}),$ 
where $\mathrm{cos}(\cdot)$ denotes the cosine similarity function, and $e_t$ represents the normalized embedding of query $q_t$ (i.e., $|e_t| = 1$). 

\medskip\noindent
\textbf{Original Question Similarity (Sim).}
As a guardrail, we define a query similarity signal 
to ensure that novelty remains aligned with the topic of the user’s question and prevents query drift.
This guardrail signal is defined as $\mathrm{Sim}_{t} = \mathrm{cos}(e_t, e_x),$ 
where $e_x$ denotes the normalized embedding of the user query $x$. When this similarity becomes too low, it serves as a warning indicator that exploration may be drifting off-topic.


\renewcommand{\arraystretch}{1.4}
\begin{table}[t]
  \centering
  \small
  \setlength{\tabcolsep}{2.4pt}
  \begin{tabular}{l|ccc|cc|cc}
    \hline
    \multirow{2}{*}{\textbf{Agent}} &
    \multicolumn{3}{c|}{BrowseCompP.} &\multicolumn{2}{c|}{NeuCLIR} & \multicolumn{2}{c}{LegalSearch} \\
    \cline{2-8}
    & R & Acc & \#S$\downarrow$  & R & \#S$\downarrow$  & R & \#S $\downarrow$ \\\hline\hline
    CPM-Report       & 18.6 & 11.7 & 14.4 & 44.5 & 20.6 & 31.8 & 19.4 \\
    + \proposedmethod & \textbf{30.5\textsuperscript{\dag}} & \textbf{12.0} & 18.2 & \textbf{48.7} & 22.2  & \textbf{40.8\textsuperscript{\dag}} & 21.6 \\
    \hline
    GPT-oss-20b      & 43.0 & 23.0 & 28.3 & 41.8 & 12.2 & 40.5 & 10.2 \\
    + \proposedmethod & \textbf{50.7\textsuperscript{\dag}} & \textbf{26.4\textsuperscript{\dag}} & 27.0 & \textbf{43.7} & 14.0 & \textbf{43.8} &10.8 \\
    \hline
    CPM-Explore      & 40.1 & \textbf{30.5} & 29.6 & \textbf{52.1} & 25.5 & 44.2 & 23.8 \\
    + \proposedmethod & \textbf{48.0\textsuperscript{\dag}} & 29.7 & 21.7 & 42.5\textsuperscript{\dag} & 12.5 & \textbf{45.2} & 12.2 \\
    \hline
    GLM-4.7          & 55.9 & 32.0 & 50.3 & 43.9 & 12.1 & 45.0 & 19.8 \\
    + \proposedmethod & \textbf{59.2\textsuperscript{\dag}} & \textbf{34.9\textsuperscript{\dag}} & 36.8 & \textbf{46.8} & 17.1 & \textbf{50.0} & 18.6 \\
    \hline
    ReAct            & 60.9 & 35.7 & 35.3 & 35.9 & 5.6 & 42.6 & 8.7 \\
    + \proposedmethod & \textbf{62.6} & \textbf{39.1} & 25.4 & \textbf{39.1} & 5.7 & \textbf{47.5\textsuperscript{\dag}} & 8.4 \\
    \hline
    WebWeaver        & 64.9 & 47.8 & 75.0 & 49.0 & 19.2 & 47.5 & 19.9 \\
    + \proposedmethod &  \textbf{\underline{67.2}\textsuperscript{\dag}} & \textbf{\underline{51.3}\textsuperscript{\dag}} & 50.7 & \textbf{50.0} & 20.3 & \textbf{51.6} & 22.1 \\
    \hline
    Tongyi-DR        & 48.4 & 37.9 & 77.8 & 49.1 & 62.1 & 52.2 & 53.8 \\
    + \proposedmethod & \textbf{59.0\textsuperscript{\dag}} & \textbf{47.9\textsuperscript{\dag}} & 33.6 & \textbf{\underline{55.6}\textsuperscript{\dag}} & 34.4 & \textbf{\underline{56.5}} & 22.8 \\
    \hline\hline
  \end{tabular}
  \caption{Performance of DRAs without and with the \proposedmethod. \textbf{R}: recall, \textbf{Acc}: accuracy, and \textbf{\#S}: number of search calls. The numbers in \textbf{bold} and with \underline{underline} show the best results for an agent and overall, respectively. Significant differences are marked with \textsuperscript{\dag}. Overall trend: enhancing DRAs with \proposedmethod\ improves the performance of the agents and simultaneously decreases the number of search calls.}
  \label{tab:controller_performance}
\end{table}

\subsection{Agent Controller}~\label{subsec:controller}
We now explain Retrieval-Aware Agent Controller (\proposedmethod) as an approach to incorporate the introduced signals in a given DRA. The core idea of \proposedmethod\ is that, the controller monitors the agent’s interaction history using the proposed signals and adjusts the process of the agent when needed. At each iteration $t$, the controller computes the retrieval-aware signals (i.e., $\mathrm{Nov}_{t}, \mathrm{Cov}_{t}, \mathrm{Div}_{t},\mathrm{and~Sim}_{t}$), and suggests an action from a predefined set based on the previous turns and their corresponding signals. The action set is defined as 
$A = \{ \textit{continue}, \textit{intervene}, \textit{stop} \}, $ and explained in what follows. This overall architecture of \proposedmethod\ is depicted in Figure~\ref{fig:example} in Appendix~\ref{sec:app:exp}. 

\medskip\noindent
\textbf{Continue.}
This action simply allows the agent to proceed with no intervention.

\medskip\noindent
\textbf{Intervene.}
As discussed in previous studies~\cite{Soudani26Uncertainty, Li25RASPberry}, when the trajectory becomes stale but still recoverable, an external nudge (\eg critical reasoning) can redirect the search toward a more informative direction. Upon invoking \emph{Intervene}, the controller calls a critical re-thinker, which (1) critically assesses why the current search trajectory is stagnating, and (2) generates a substantially different and potentially creative search query to help escape the loop.
More specifically, 
\new{the critical re-thinker model} takes $\mathcal{H}_{t}$ as input and generates a 
\new{critical reasoning thought} $\tau^{cr}_{t+1}$ together with a new search query $q^{cr}_{t+1}$. The corresponding document set $D^{cr}_{t+1}$ is then retrieved, and the resulting triplet is appended to the interaction history:
$$
\mathcal{H}_{t+1} = \mathcal{H}_{t} \oplus \big( \tau^{cr}_{t+1}, q^{cr}_{t+1}, D^{cr}_{t+1} \big).
$$

\medskip\noindent
\textbf{Stop.}
This action is selected when the trajectory is exhausted and further iterations are unlikely to yield new information~\citep{park2025stop, Hu25MCTSRAG}. Given $\mathcal{H}_{t}$, the agent generates a final reasoning step $\tau^{\prime}_T$, which leads to the final answer $r_T$, defined as: $\tau^{\prime}_T, r_T \sim \pi(\cdot \mid f(\mathcal{H}_{t})).$

\subsection{Experiments and Results}
We follow the same experimentation setup as explained in Section~\ref{sec:analysis_results}. To calculate $\mathrm{Cov}$, the required criteria are provided in \textsc{BrowseComp-Plus}. For \textsc{NeuCLIR}~\citep{lawrie2025neuclir} and \textsc{LegalSearch}, we implement an LLM-based component to create these criteria and further update them (details provided in Appendix~\ref{sec:app:exp}). For the signals which require cosine similarity, we use \verb|Qwen3-Embedding-4B| as the encoder. \new{We use Claude Sonnet 4.5 as the backbone LLM for the criteria coverage updater, controller, and critical re-thinker used for intervention actions. 
} Besides the overall accuracy, we report (overall) recall, which measures the proportion of relevant documents returned across \emph{all} search calls for a query~\cite{meng2026revisiting}. Statistical significance is reported using $t$-test for recall and McNemar test for accuracy ($p < 0.05$). Appendix~\ref{sec:app:exp} provides the details of the experimentation setup and applied prompts.

\medskip
\noindent
\textbf{(RQ3) What is the effect of \proposedmethod\ in mitigating reasoning stagnation? }
The evaluation results of the agents with and without \proposedmethod\ are reported in Table~\ref{tab:controller_performance}. Since \textsc{BrowseComp-Plus} contains gold criteria, all four signals of $\mathrm{Cov}_{t}$, $\mathrm{Nov}_{t}$, $\mathrm{Div}_{t}$, $\mathrm{Sim}_{t}$. On \textsc{NeuCLIR} and \textsc{LegalSearch}, \proposedmethod\ uses the same signals except $\mathrm{Cov}_{t}$.

Focusing first on the performance metrics, namely recall on all datasets and accuracy on \textsc{BrowseComp-Plus}. The results show that, across all agents after applying \proposedmethod, the performance metrics either improve or remain on par. 
From an efficiency point of view, we observe a reduction in the number of search calls / iterations on \textsc{BrowseComp-Plus} for all agents except CPM-Report. Interestingly, the increase in the number of search calls for CPM-Report leads to a significant improvement in both performance metrics, as the original agent (in contrast to the others) inherently tends to terminate too early, corrected by \proposedmethod. 

A similar pattern is observed in \textsc{NeuCLIR} and \textsc{LegalSearch}. On these datasets, applying \proposedmethod\ mostly reduces the number of search calls. In cases where the number of iterations increases, we consistently observe an increase in performance, showing the adaptive behavior of \proposedmethod\ to iterate more when more evidence is needed. 

On all datasets, the best overall performance is achieved by applying \proposedmethod\ to the previous SOTA agents. In particular, on \textsc{BrowseComp-Plus} WebWeaver+\proposedmethod\ achieves the best results by significantly improving WebWeaver on both recall and accuracy, and simultaneously decreasing the number of its search calls by 33\%. On \textsc{NeuCLIR}, Tongyi-DR+\proposedmethod\ achieves the best overall recall and significantly improves its base agent, while decreasing the average search calls by 45\%. The same pattern is observed on \textsc{LegalSearch} with an improvement in recall followed by 58\% decrease in the number of iterations.
\new{Although the controller introduces additional latency due to its multiple LLM calls, the total overhead of the RAAC controller per query is only approximately 10\% of the latency required for the agent's reasoning and retrieval process. This overhead is small compared to the average 33\% reduction in search calls, as the controller components execute well-defined operations and produce significantly shorter outputs than the agent's reasoning trajectories.}

\renewcommand{\arraystretch}{1.5}
\begin{table}[t]
  \centering
  \small
  \shrink
  \setlength{\tabcolsep}{2.4pt}
  \begin{tabular}{ll|ccc|cc|cc}
    \hline
    \multirow{2}{*}{\textbf{Ag.}} & \multirow{2}{*}{\textbf{Signals}} &
    \multicolumn{3}{c|}{BrowseCompP.} &\multicolumn{2}{c|}{NeuCLIR} & \multicolumn{2}{c}{LegalSearch} \\
    \cline{3-9}
    && R & Acc & \#S$\downarrow$  & R & \#S$\downarrow$  & R & \#S $\downarrow$ \\
    \hline\hline
    \multirow{3}{*}{\rotatebox[origin=c]{90}{OSS}} &
    $N$-$C$-$D$   & \cellcolor{gray!20}\textbf{50.7} & \cellcolor{gray!20}\textbf{26.4} & \cellcolor{gray!20}27.0 & 39.3 & 11.6 & 40.2 & 10.0 \\
    & $N$-$D$   & 48.5\textsuperscript{\dag} & 24.5\textsuperscript{\dag} & 25.5 & \cellcolor{gray!20}\textbf{43.7} & \cellcolor{gray!20}14.0 & \cellcolor{gray!20}\textbf{43.8} & \cellcolor{gray!20}10.8 \\
    & $C$-$D$   & 41.4\textsuperscript{\dag} & 17.2\textsuperscript{\dag} & 12.3 & 39.3 & 10.4 & 40.7 & 8.9  \\
    \hline
    \multirow{3}{*}{\rotatebox[origin=c]{90}{GLM}} &
      $N$-$C$-$D$ & \cellcolor{gray!20}\textbf{59.2} & \cellcolor{gray!20}\textbf{34.9} & \cellcolor{gray!20}36.8 & 44.1 & 12.0 & 47.2 & 13.9 \\
    & $N$-$D$   & 56.3\textsuperscript{\dag} & 33.9 & 35.3 & \cellcolor{gray!20}\textbf{46.8} & \cellcolor{gray!20}17.1 & \cellcolor{gray!20}\textbf{50.0} & \cellcolor{gray!20}18.6 \\
    & $C$-$D$   & 45.1\textsuperscript{\dag} & 26.9\textsuperscript{\dag} & 14.7 & 41.1\textsuperscript{\dag} & 10.0 & 44.4\textsuperscript{\dag} & 11.2 \\
    \hline
    \multirow{3}{*}{\rotatebox[origin=c]{90}{WebWeaver}} &
     $N$-$C$-$D$  & \cellcolor{gray!20}\textbf{67.2} & \cellcolor{gray!20}51.3 & \cellcolor{gray!20}50.7 & 47.9 & 19.8 & 50.1 & 19.1 \\
    & $N$-$D$   & 65.7 & \textbf{51.8} & 42.1 & \cellcolor{gray!20}\textbf{50.0} & \cellcolor{gray!20}20.3 & \cellcolor{gray!20}\textbf{51.6} & \cellcolor{gray!20}22.1 \\
    & $C$-$D$   & 56.7\textsuperscript{\dag} & 43.2\textsuperscript{\dag} & 25.6 & 46.5 & 16.9 & 46.2 & 16.4 \\
    \hline
    \multirow{3}{*}{\rotatebox[origin=c]{90}{Tongyi}} &
     $N$-$C$-$D$  & \cellcolor{gray!20}\textbf{59.0} & \cellcolor{gray!20}47.9 & \cellcolor{gray!20}33.6 & 47.9\textsuperscript{\dag} & 21.1 & 49.7\textsuperscript{\dag} & 14.4 \\
    & $N$-$D$   & 56.9 & \textbf{49.1} & 31.3 & \cellcolor{gray!20}\textbf{55.6} & \cellcolor{gray!20}34.4 & \cellcolor{gray!20}\textbf{56.5} & \cellcolor{gray!20}22.8 \\
    & $C$-$D$   & 48.1\textsuperscript{\dag} & 45.6\textsuperscript{\dag} & 18.1 & 47.8\textsuperscript{\dag} & 16.9 & 43.6\textsuperscript{\dag} & 12.1 \\
    \hline
    \hline
  \end{tabular}
  \miniskip
  \caption{Evaluation of the agents with RAAC under different sets of signals. $C$, $N$, and $D$ denote Criteria Coverage ($\mathrm{Cov}$), Document Novelty ($\mathrm{Nov}$), and Query Diversity ($\mathrm{Div}$), respectively. Significant differences are compared against the \colorbox{gray!20}{gray row} in each block and marked with \textsuperscript{\dag}.}
  \label{tab:ablation_study}
  \shrink
\end{table}

\begin{figure}[t]
\shrink
  \centering
  \includegraphics[width=0.48\textwidth]{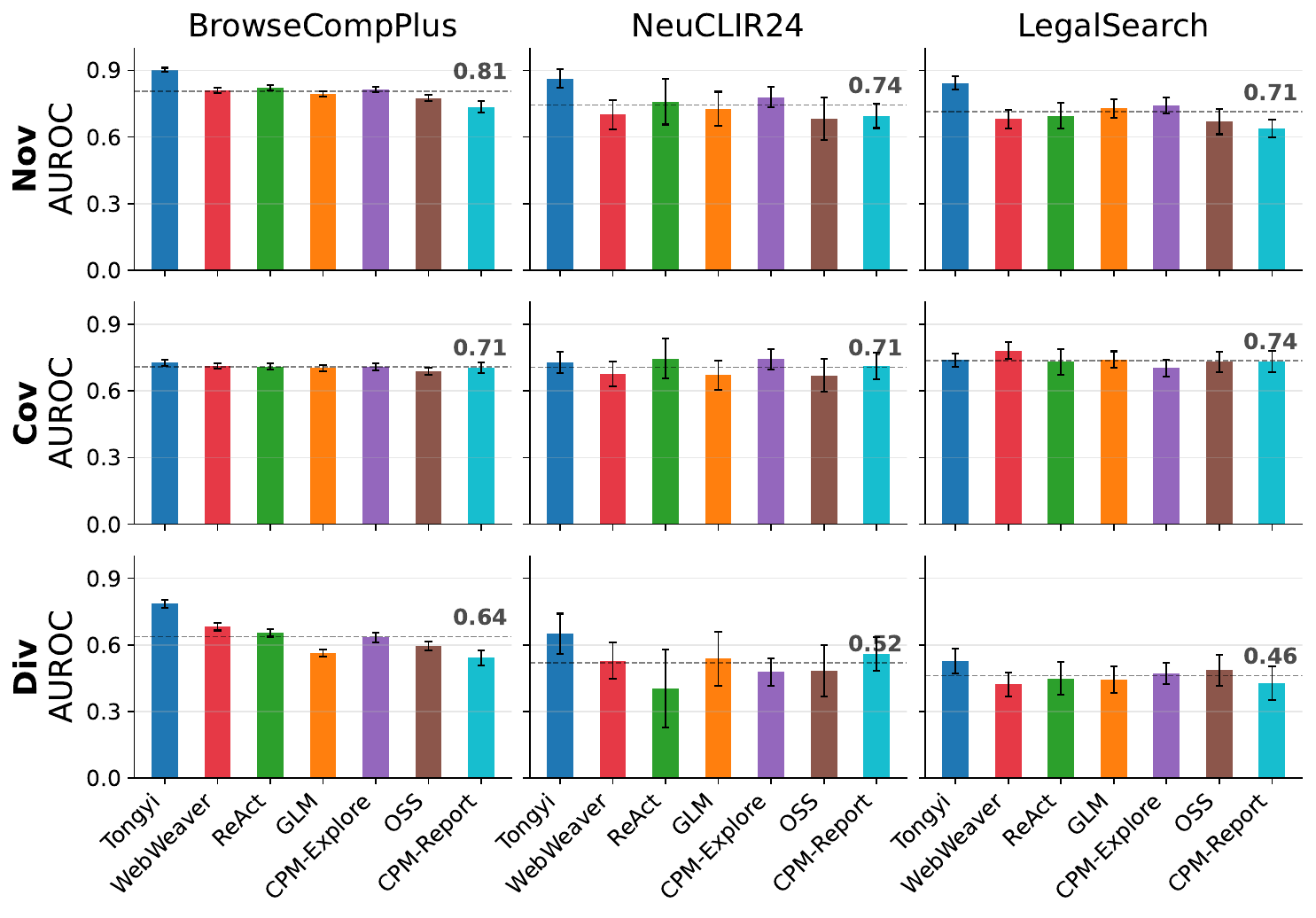}
  \shrink
  \caption{The correlation between unsupervised signals and New-item Recall, measured using AUROC. 
  }
  \label{fig:correlation_signals_auroc}
  \shrink
\end{figure}


To summarize, applying \proposedmethod\ to the DRAs under the study in most cases reduces unnecessary iterations through targeted interventions, and in some cases increases the iterations to accumulate more evidence. The results demonstrate that the importance of signal-driven controller for DRAs to improve efficiency and effectiveness and to support with robust adaptivity across different trajectories.

\medskip 
\noindent
\textbf{(RQ4) What is the effect of each signal on controlling DRAs behavior? } 
Table~\ref{tab:ablation_study} presents the results of an ablation study of the controller under different signal combinations. In this table, we denote $\mathrm{Cov}$ as $C$, $\mathrm{Nov}$ as $N$, and $\mathrm{Div}$ as $D$. We keep $\mathrm{Sim}$ in all setups, as it is defined as the guardrail signal. 
We report results for four agents due to context limitations. Full results are in Table~\ref{tab:dra_performance_all}.

On \textsc{BrowseComp-Plus}, incorporating Cov improves performance, whereas on \textsc{NeuCLIR} and \textsc{LegalSearch}, better results are achieved without it. This is due to the presence of human-generated criteria in \textsc{BrowseComp-Plus} queries and suggests that LLM-generated criteria are useful only when human-generated criteria are not available in the query.
Across all datasets, removing $\mathrm{Nov}$ as the primary signal reduces recall, indicating that $\mathrm{Nov}$ is the most effective choice for the primary signal. 

To further analyze the informativeness of each signal (independent of its integration in \proposedmethod), we compute the correlation between each signal and recall.
We define \textbf{\textit{New-item Recall}} as the fraction of new relevant documents retrieved at iteration $t$ that have not been retrieved in previous iterations. Formally,
$$\text{NiR}_t = \frac{\left| (D_t \cap D_g) \setminus D_{1:t-1} \right|}{|D_g|}.$$
We argue that the correlation between unsupervised signals and \textit{New-item Recall} serves as a reliable proxy for measuring the informativeness of these signals. 
Therefore, we evaluate how well each signal discriminates between productive iterations ($\text{NiR}_t > 0$) and non-productive iterations ($\text{NiR}_t = 0$).
Because \new{$\text{NiR}$} is \emph{zero-inflated}, with most iterations failing to retrieve any new relevant documents, we report AUROC, as it quantifies how well a signal distinguishes between productive and non-productive iterations, making it particularly appropriate for sparse binary-like outcomes~\citep{McDermott2024AUROC, soudani-2025-uncertainty}.

Figure~\ref{fig:correlation_signals_auroc} presents the correlation between \new{$\text{NiR}$} and each signal (apart from the guardrail Sim$_t$ signal) across all setups. On average, $\mathrm{Nov}_t$ exhibits the highest correlation, followed by $\mathrm{Cov}_t$, while $\mathrm{Div}_t$ shows the lowest correlation. These results are largely consistent with the roles in the controller. 
Therefore, the correlation with \new{$\text{NiR}_t$} serves as a valid proxy for measuring the effectiveness of the unsupervised signals, highlighting the importance of $\mathrm{Nov}_t$ and $\mathrm{Cov}_t$ as key signals to control stagnant reasoning in DRAs.

\begin{figure}[t]
  \centering
  \includegraphics[width=0.48\textwidth]{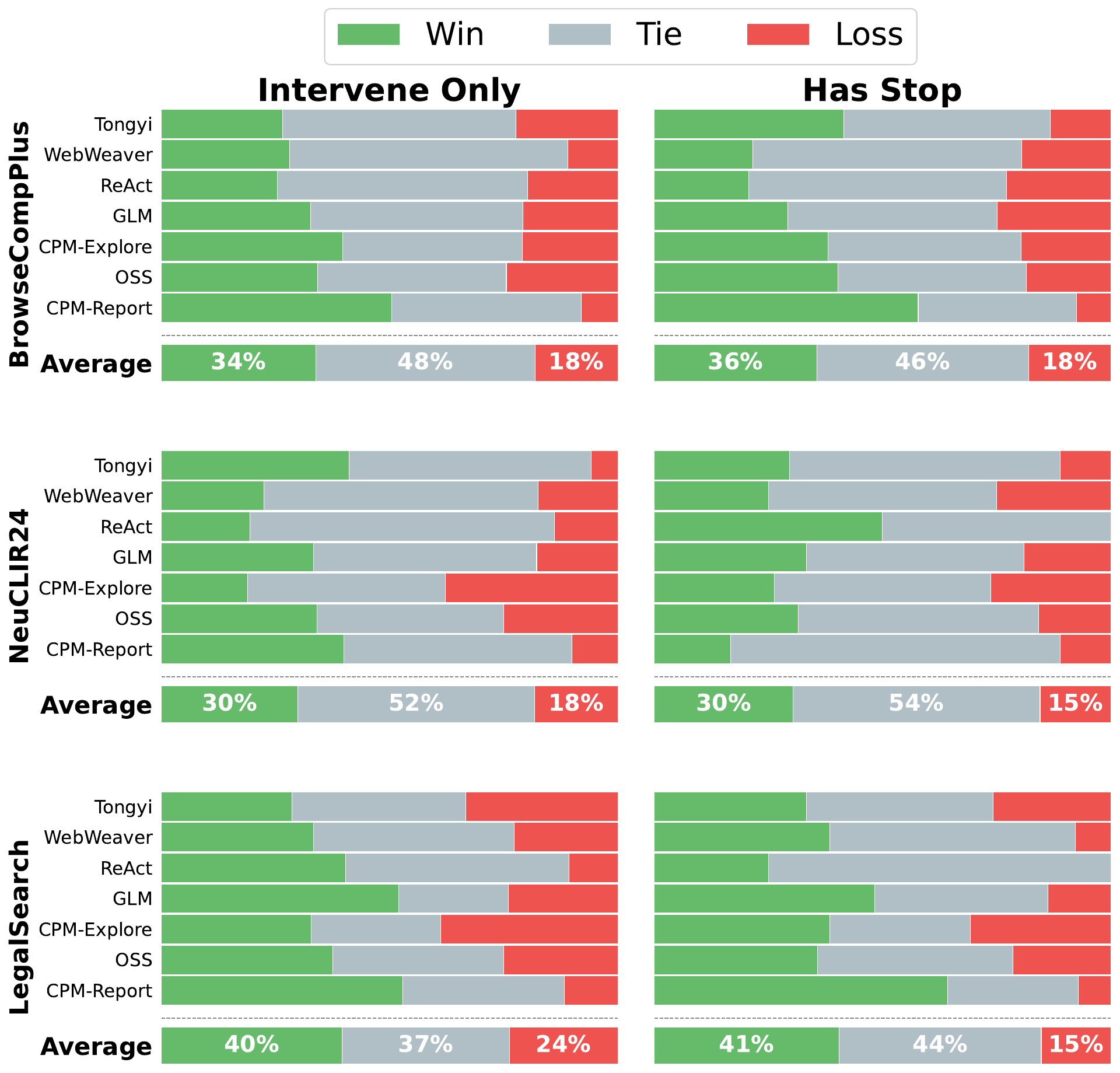}
  \caption{Evaluation of the impact of the controller’s actions measured by recall.}
  \label{fig:controller_effect}
\end{figure}

\medskip 
\noindent
\textbf{(RQ5) What is the effect of \proposedmethod\ actions on steering the trajectory toward an optimal path?}
We analyze the effect of the two active controller actions, intervene and stop. 
We partition the queries into two groups based on the controller's behavior during the trajectory:
\emph{Has~Stop}, which contains queries for which the controller issued the Stop action, and 
\emph{Intervene~Only}, which contains queries where the controller \textit{only} issued Intervene action and never stopped the agent early.
Within each group, we compute the final number of retrieved gold documents with and without \proposedmethod\ and categorize queries to \emph{win} (more gold documents with controller), \emph{loss} (fewer gold documents), or a \emph{tie} (equal count) group.

Figure~\ref{fig:controller_effect} presents the win/tie/loss distributions across all agents and datasets. In the \textit{Has~Stop} group, wins consistently outnumber losses, with average win rates of 35\% vs.\ 18\% on \textsc{BrowseComp-Plus}, 30\% vs.\ 15\% on \textsc{NeuCLIR}, and 41\% vs.\ 15\% on \textsc{LegalSearch}. The \textit{Intervene~Only} group shows a similar positive trend (33\% vs.\ 18\%, 30\% vs.\ 18\%, and 40\% vs.\ 24\%, respectively). These results suggest that both controller actions are effective: intervention helps redirect trajectories toward higher-recall paths, while stopping does not prematurely terminate productive iterations.
\new{Appendix~\ref{sec:app:case_study} provides additional case studies that further validate these findings and analyze potential false-stagnation scenarios.}

We further analyze how samples are distributed across groups. On \textsc{BrowseComp-Plus}, 61.5\% of query--agent pairs fall into the \textit{Has~Stop} group and 34.6\% into the \textit{Intervene~Only} group, indicating that the controller actively intervenes in the majority of trajectories. In contrast, on \textsc{NeuCLIR} and \textsc{LegalSearch}, the pattern reverses: \textit{Intervene~Only} dominates (55.4\% and 54.8\%, respectively). This reflects the shorter trajectories in these datasets, where the controller more often redirects behavior than terminates it.

\section{Discussion and Conclusions}~\label{sec:conclusions}
In this paper, we analyze the reasoning behavior of DRAs through intermediate recall and accuracy, revealing that these agents suffer from stagnant reasoning, where many iterations yield little to no meaningful performance improvement.
We then argue that agents should be informed about the progress of their reasoning trajectory, which can be achieved through trajectory-based signals. To this end, we propose four unsupervised signals: document novelty, criteria coverage, search query diversity, and original query similarity. Based on these signals, we design a controller that selects an action from a predefined set (continue, intervene, or stop) in order to redirect the reasoning trajectory toward a more optimal path.
Our experimental results across seven agents and three datasets show that the controller consistently improves effectiveness while reducing latency, validating the usefulness of the proposed signals.
A promising direction for future work is to train agents that are inherently aware of unsupervised trajectory signals during task completion.

\section*{Limitations}

\noindent\textbf{Prompt-based Controller.}
Our controller (Section~\ref{subsec:controller}) relies on a hand-crafted prompt to map trajectory signals into one of three discrete actions: continue, intervene, or stop. While this design is effective and requires no training data, it applies fixed weights to these signals uniformly across all queries, rather than adapting them to the characteristics of individual queries.
A learned controller could instead adaptively weight the signals, capture dependencies across successive observations, and potentially generalize better to diverse query types. We leave the exploration of learned controllers, including supervised and reinforcement learning-based approaches, to future work.

\medskip 
\noindent\textbf{Comprehensiveness of the Signals.}
The \proposedcomponent\ leverages unsupervised trajectory signals to estimate task completion progress.
While we introduce four signals and analyze the contribution of each, we are interested in identifying additional signals to make the decision-making process more informed.
For example, signals derived from document content, retrieval confidence scores, or inter-document redundancy may provide complementary information.
Future work could explore a broader range of trajectory signals and investigate their interactions.

\medskip 
\noindent\textbf{Comprehensiveness of the Actions.}
Our analysis identifies two recurring failure modes in DRAs: failing to stop at the optimal point and becoming trapped in a narrow search space without exploring alternative perspectives.
To address these issues, we introduce two controller actions: stop and intervene (through critical re-thinking), alongside a continue action that preserves the agent’s default behavior.
Although these actions target the most prominent shortcomings observed in our analysis, DRAs are a rapidly evolving paradigm and may exhibit additional failure modes as they are applied to a broader range of tasks.
Future work could explore these limitations further and introduce additional actions to address them.

\section*{Acknowledgments}
This work was supported in part by the project LESSEN with project number NWA.1389.20.183 of the research program NWA ORC 2020/21 which is (partly) financed by the Dutch Research Council (NWO). This work was carried out during the internship of the first author at Thomson Reuters Labs.

\bibliography{custom}

@inproceedings{Noise2024Florin,
  author       = {Florin Cuconasu and
                  Giovanni Trappolini and
                  Federico Siciliano and
                  Simone Filice and
                  Cesare Campagnano and
                  Yoelle Maarek and
                  Nicola Tonellotto and
                  Fabrizio Silvestri},
  title        = {The Power of Noise: Redefining Retrieval for {RAG} Systems},
  booktitle    = {Proceedings of the 47th International {ACM} {SIGIR} Conference on Research and Development in Information Retrieval, {SIGIR}},
  pages        = {719--729},
  year         = {2024},
  url = {https://doi.org/10.1145/3626772.3657834}
}

@inproceedings{mallen23popqa,
    title = "When Not to Trust Language Models: Investigating Effectiveness of Parametric and Non-Parametric Memories",
    author = "Mallen, Alex  and
      Asai, Akari  and
      Zhong, Victor  and
      Das, Rajarshi  and
      Khashabi, Daniel  and
      Hajishirzi, Hannaneh",
    booktitle = "Proceedings of the 61st Annual Meeting of the Association for Computational Linguistics (Volume 1: Long Papers)",
    month = jul,
    year = "2023",
    pages = "9802--9822",
        url = "https://aclanthology.org/2023.acl-long.546/",

}

@inproceedings{Soudani24FTvsRAG,
  author       = {Heydar Soudani and
                  Evangelos Kanoulas and
                  Faegheh Hasibi},
  title        = {Fine Tuning vs. Retrieval Augmented Generation for Less Popular Knowledge},
  booktitle    = {Proceedings of the 2024 Annual International {ACM} {SIGIR} Conference on Research and Development in Information Retrieval in the Asia Pacific Region, {SIGIR-AP} 2024},
  pages        = {12--22},
  year         = {2024},
  url = {https://doi.org/10.1145/3673791.3698415},

}

@inproceedings{trivedi2023ircot,
    title = "Interleaving Retrieval with Chain-of-Thought Reasoning for Knowledge-Intensive Multi-Step Questions",
    author = "Trivedi, Harsh  and
      Balasubramanian, Niranjan  and
      Khot, Tushar  and
      Sabharwal, Ashish",
    booktitle = "Proceedings of the 61st Annual Meeting of the Association for Computational Linguistics (Volume 1: Long Papers)",
    year = "2023",
    pages = "10014--10037",
}

@inproceedings{jiang2023flare,
    title = "Active Retrieval Augmented Generation",
    author = "Jiang, Zhengbao  and
      Xu, Frank  and
      Gao, Luyu  and
      Sun, Zhiqing  and
      Liu, Qian  and
      Dwivedi-Yu, Jane  and
      Yang, Yiming  and
      Callan, Jamie  and
      Neubig, Graham",
    booktitle = "Proceedings of the 2023 Conference on Empirical Methods in Natural Language Processing (EMNLP)",
    month = dec,
    year = "2023",
    pages = "7969--7992",
        url = "https://aclanthology.org/2023.emnlp-main.495/"

}

@inproceedings{su2024dragin,
    title = "{DRAGIN}: Dynamic Retrieval Augmented Generation based on the Real-time Information Needs of Large Language Models",
    author = "Su, Weihang  and
      Tang, Yichen  and
      Ai, Qingyao  and
      Wu, Zhijing  and
      Liu, Yiqun",
    booktitle = "Proceedings of the 62nd Annual Meeting of the Association for Computational Linguistics (Volume 1: Long Papers)",
    month = aug,
    year = "2024",
    pages = "12991--13013",
}

@inproceedings{press2023selfask,
    title = "Measuring and Narrowing the Compositionality Gap in Language Models",
    author = "Press, Ofir  and
      Zhang, Muru  and
      Min, Sewon  and
      Schmidt, Ludwig  and
      Smith, Noah  and
      Lewis, Mike",
    booktitle = "Findings of the Association for Computational Linguistics: EMNLP",
    month = dec,
    year = "2023",
}

@inproceedings{Yao23ReAct,
  author       = {Shunyu Yao and
                  Jeffrey Zhao and
                  Dian Yu and
                  Nan Du and
                  Izhak Shafran and
                  Karthik R. Narasimhan and
                  Yuan Cao},
  title        = {ReAct: Synergizing Reasoning and Acting in Language Models},
  booktitle    = {The Eleventh International Conference on Learning Representations, {ICLR}},
  year         = {2023},
}

@inproceedings{li25searcho1,
  author       = {Xiaoxi Li and
                  Guanting Dong and
                  Jiajie Jin and
                  Yuyao Zhang and
                  Yujia Zhou and
                  Yutao Zhu and
                  Peitian Zhang and
                  Zhicheng Dou},
  title        = {Search-o1: Agentic Search-Enhanced Large Reasoning Models},
  booktitle    = {Proceedings of the 2025 Conference on Empirical Methods in Natural
                  Language Processing, {EMNLP}},
  pages        = {5420--5438},
  publisher    = {Association for Computational Linguistics},
  year         = {2025},
      url = "https://aclanthology.org/2025.emnlp-main.276/",

}

@inproceedings{
Jin25SearchR1,
title={Search-R1: Training {LLM}s to Reason and Leverage Search Engines with Reinforcement Learning},
author={Bowen Jin and Hansi Zeng and Zhenrui Yue and Jinsung Yoon and Sercan O Arik and Dong Wang and Hamed Zamani and Jiawei Han},
booktitle={Second Conference on Language Modeling},
year={2025},
url={https://openreview.net/forum?id=Rwhi91ideu}
}

@article{meng2026revisiting,
    title={Revisiting Text Ranking in Deep Research},
    author={Chuan Meng and Litu Ou and Sean MacAvaney and Jeff Dalton},
    year={2026},
    journal={arXiv preprint arXiv:2602.21456},
    url={https://arxiv.org/abs/2602.21456},
}

@article{chen2026agentir,
    title={{AgentIR}: Reasoning-Aware Retrieval for Deep Research Agents},
    author={Zijian Chen and Xueguang Ma and Shengyao Zhuang and Jimmy Lin and Akari Asai and Victor Zhong},
    year={2026},
    journal={arXiv preprint arXiv:2603.04384},
    url={https://arxiv.org/abs/2603.04384},
}

@article{sharifymoghaddam2026rerank,
    title={Rerank Before You Reason: Analyzing Reranking Tradeoffs through Effective Token Cost in Deep Search Agents},
    author={Sahel Sharifymoghaddam and Jimmy Lin},
    year={2026},
    journal={arXiv preprint arXiv:2601.14224},
    url={https://arxiv.org/abs/2601.14224},
}

@article{glm2025arc,
    title={{GLM-4.5}: Agentic, Reasoning, and Coding ({ARC}) Foundation Models},
    author={{GLM Team}},
    year={2025},
    journal={arXiv preprint arXiv:2508.06471},
    url={https://arxiv.org/abs/2508.06471},
}

@article{openai2025gptoss,
    title={gpt-oss-120b \& gpt-oss-20b Model Card},
    author={{OpenAI}},
    year={2025},
    journal={arXiv preprint arXiv:2508.10925},
    url={https://arxiv.org/abs/2508.10925},
}

@article{tongyi2025deepresearch,
    title={Tongyi {DeepResearch} Technical Report},
    author={{Tongyi DeepResearch Team}},
    year={2025},
    journal={arXiv preprint arXiv:2510.24701},
    url={https://arxiv.org/abs/2510.24701},
}

@article{li2025webweaver,
    title={{WebWeaver}: Structuring Web-Scale Evidence with Dynamic Outlines for Open-Ended Deep Research},
    author={Zijian Li and Xin Guan and Bo Zhang and Shen Huang and Houquan Zhou and Shaopeng Lai and Ming Yan and Yong Jiang and Pengjun Xie and Fei Huang and Jun Zhang and Jingren Zhou},
    year={2025},
    journal={arXiv preprint arXiv:2509.13312},
    url={https://arxiv.org/abs/2509.13312},
}

@article{li2026agentcpm,
    title={{AgentCPM-Report}: Interleaving Drafting and Deepening for Open-Ended Deep Research},
    author={Yishan Li and Wentong Chen and Yukun Yan and Mingwei Li and Sen Mei and Xiaorong Wang and Kunpeng Liu and Xin Cong and Shuo Wang and Zhong Zhang and Yaxi Lu and Zhenghao Liu and Yankai Lin and Zhiyuan Liu and Maosong Sun},
    year={2026},
    journal={arXiv preprint arXiv:2602.06540},
    url={https://arxiv.org/abs/2602.06540},
}

@article{lawrie2025neuclir,
    title={Overview of the {TREC} 2024 {NeuCLIR} Track},
    author={Dawn Lawrie and Sean MacAvaney and James Mayfield and Paul McNamee and Douglas W. Oard and Luca Soldaini and Eugene Yang},
    year={2025},
    journal={arXiv preprint arXiv:2509.14355},
    url={https://arxiv.org/abs/2509.14355},
}

@article{Support25Thakur,
  author       = {Nandan Thakur and
                  Ronak Pradeep and
                  Shivani Upadhyay and
                  Daniel Campos and
                  Nick Craswell and
                  Jimmy Lin},
  title        = {Support Evaluation for the {TREC} 2024 {RAG} Track: Comparing Human
                  versus {LLM} Judges},
  journal      = {CoRR},
  volume       = {abs/2504.15205},
  year         = {2025},
  url          = {https://doi.org/10.48550/arXiv.2504.15205},
  doi          = {10.48550/ARXIV.2504.15205},
  eprinttype   = {arXiv},
  eprint       = {2504.15205},
  bibsource    = {dblp computer science bibliography, https://dblp.org}
}

@inproceedings{Enabling23Gao,
  author       = {Tianyu Gao and
                  Howard Yen and
                  Jiatong Yu and
                  Danqi Chen},
  title        = {Enabling Large Language Models to Generate Text with Citations},
  booktitle    = {Proceedings of the 2023 Conference on Empirical Methods in Natural
                  Language Processing, {EMNLP}},
  pages        = {6465--6488},
  publisher    = {Association for Computational Linguistics},
  year         = {2023},
      url = "https://aclanthology.org/2023.emnlp-main.398/",

}

@article{chen2025browsecompplus,
    title={{BrowseComp-Plus}: A More Fair and Transparent Evaluation Benchmark of Deep-Research Agent},
    author={Zijian Chen and Xueguang Ma and Shengyao Zhuang and Ping Nie and Kai Zou and Andrew Liu and Joshua Green and Kshama Patel and Ruoxi Meng and Mingyi Su and Sahel Sharifymoghaddam and Yanxi Li and Haoran Hong and Xinyu Shi and Xuye Liu and Nandan Thakur and Crystina Zhang and Luyu Gao and Wenhu Chen and Jimmy Lin},
    year={2025},
    journal={arXiv preprint arXiv:2508.06600},
    url={https://arxiv.org/abs/2508.06600},
}

@article{qwen2025embedding,
    title={{Qwen3 Embedding}: Advancing Text Embedding and Reranking Through Foundation Models},
    author={{Qwen Team}},
    year={2025},
    journal={arXiv preprint arXiv:2506.05176},
    url={https://arxiv.org/abs/2506.05176},
}

@article{korikov2025batched,
    title={Batched Self-Consistency Improves {LLM} Relevance Assessment and Ranking},
    author={Anton Korikov and Pan Du and Scott Sanner and Navid Rekabsaz},
    year={2025},
    journal={arXiv preprint arXiv:2505.12570},
    url={https://arxiv.org/abs/2505.12570},
}

@article{Li25Tongyi,
  author       = {Baixuan Li and
                  Bo Zhang and
                  Dingchu Zhang and
                  Fei Huang and
                  Guangyu Li and
                  Guoxin Chen and
                  Huifeng Yin and
                  Jialong Wu and
                  Jingren Zhou and
                  Kuan Li and
                  Liangcai Su and
                  Litu Ou and
                  Liwen Zhang and
                  Pengjun Xie and
                  Rui Ye and
                  Wenbiao Yin and
                  Xinmiao Yu and
                  Xinyu Wang and
                  Xixi Wu and
                  Xuanzhong Chen and
                  Yida Zhao and
                  Zhen Zhang and
                  Zhengwei Tao and
                  Zhongwang Zhang and
                  Zile Qiao and
                  Chenxi Wang and
                  Donglei Yu and
                  Gang Fu and
                  Haiyang Shen and
                  Jiayin Yang and
                  Jun Lin and
                  Junkai Zhang and
                  Kui Zeng and
                  Li Yang and
                  Hailong Yin and
                  Maojia Song and
                  Ming Yan and
                  Peng Xia and
                  Qian Xiao and
                  Rui Min and
                  Ruixue Ding and
                  Runnan Fang and
                  Shaowei Chen and
                  Shen Huang and
                  Shihang Wang and
                  Shihao Cai and
                  Weizhou Shen and
                  Xiaobin Wang and
                  Xin Guan and
                  Xinyu Geng and
                  Yingcheng Shi and
                  Yuning Wu and
                  Zhuo Chen and
                  Zijian Li and
                  Yong Jiang},
  title        = {Tongyi DeepResearch Technical Report},
  journal      = {CoRR},
  volume       = {abs/2510.24701},
  year         = {2025},
  url          = {https://doi.org/10.48550/arXiv.2510.24701},
  doi          = {10.48550/ARXIV.2510.24701},
  eprinttype   = {arXiv},
  eprint       = {2510.24701},
  bibsource    = {dblp computer science bibliography, https://dblp.org}
}

@inproceedings{jeong2024adaptiverag,
  title={Adaptive-{RAG}: Learning to Adapt Retrieval-Augmented Large Language Models through Question Complexity},
  author={Soyeong Jeong and Jinheon Baek and Sukmin Cho and Sung Ju Hwang and Jong Park},
  booktitle={Proceedings of the 2024 Conference of the North American Chapter of the Association for Computational Linguistics, {NAACL}},
  year={2024},
      url = "https://aclanthology.org/2024.naacl-long.389/",

}

@article{park2025stop,
  title={{Stop-RAG:} Value-Based Retrieval Control for Iterative {RAG}},
  author={Park, Jaewan and Cho, Solbee and Lee, Jay-Yoon},
  journal={arXiv preprint arXiv:2510.14337},
  year={2025},
        url={https://arxiv.org/abs/2510.14337}, 

}

@article{Chen26cpmexplore,
  author       = {Haotian Chen and
                  Xin Cong and
                  Shengda Fan and
                  Yuyang Fu and
                  Ziqin Gong and
                  Yaxi Lu and
                  Yishan Li and
                  Boye Niu and
                  Chengjun Pan and
                  Zijun Song and
                  Huadong Wang and
                  Yesai Wu and
                  Yueying Wu and
                  Zihao Xie and
                  Yukun Yan and
                  Zhong Zhang and
                  Yankai Lin and
                  Zhiyuan Liu and
                  Maosong Sun},
  title        = {AgentCPM-Explore: Realizing Long-Horizon Deep Exploration for Edge-Scale
                  Agents},
  journal      = {CoRR},
  volume       = {abs/2602.06485},
  year         = {2026},
  url          = {https://doi.org/10.48550/arXiv.2602.06485},
  doi          = {10.48550/ARXIV.2602.06485},
  eprinttype   = {arXiv},
  eprint       = {2602.06485},
  bibsource    = {dblp computer science bibliography, https://dblp.org}
}

@article{li2026beyond,
  title={Beyond Semantic Similarity: Rethinking Retrieval for Agentic Search via Direct Corpus Interaction},
  author={Li, Zhuofeng and Zhang, Haoxiang and Wei, Cong and Lu, Pan and Nie, Ping and Lu, Yi and Bai, Yuyang and Feng, Shangbin and Zhu, Hangxiao and Zhong, Ming and others},
  journal={arXiv preprint arXiv:2605.05242},
  year={2026},
        url={https://arxiv.org/abs/2605.05242}, 

}

@inproceedings{min23FActScore,
  author       = {Sewon Min and
                  Kalpesh Krishna and
                  Xinxi Lyu and
                  Mike Lewis and
                  Wen{-}tau Yih and
                  Pang Wei Koh and
                  Mohit Iyyer and
                  Luke Zettlemoyer and
                  Hannaneh Hajishirzi},
  title        = {FActScore: Fine-grained Atomic Evaluation of Factual Precision in Long Form Text Generation},
  booktitle    = {Proceedings of the 2023 Conference on Empirical Methods in Natural Language Processing, {EMNLP}},
  pages        = {12076--12100},
  publisher    = {Association for Computational Linguistics},
  year         = {2023},
      url = "https://aclanthology.org/2023.emnlp-main.741/",

}

@inproceedings{Soudani26Uncertainty,
  author       = {Heydar Soudani and
                  Hamed Zamani and
                  Faegheh Hasibi},
  title        = {Uncertainty Quantification for Retrieval-Augmented Reasoning},
  year         = {2026},
  booktitle    = {Proceedings of the 49th International ACM SIGIR Conference on Research and Development in Information Retrieval}, 
  url          = {https://doi.org/10.48550/arXiv.2510.11483},
}

@inproceedings{Li25RASPberry,
  author       = {Baixuan Li and
                  Yunlong Fan and
                  Tianyi Ma and
                  Miao Gao and
                  Chuanqi Shi and
                  Zhiqiang Gao},
  title        = {RASPberry: Retrieval-Augmented Monte Carlo Tree Self-Play with Reasoning Consistency for Multi-Hop Question Answering},
  booktitle    = {Findings of the Association for Computational Linguistics, {ACL}},
  series       = {Findings of {ACL}},
  pages        = {11258--11276},
  publisher    = {Association for Computational Linguistics},
  year         = {2025},
      url = "https://aclanthology.org/2025.findings-acl.587/",
}

@inproceedings{Hu25MCTSRAG,
  author       = {Yunhai Hu and
                  Yilun Zhao and
                  Chen Zhao and
                  Arman Cohan},
  editor       = {Christos Christodoulopoulos and
                  Tanmoy Chakraborty and
                  Carolyn Rose and
                  Violet Peng},
  title        = {{MCTS-RAG:} Enhancing Retrieval-Augmented Generation with Monte Carlo
                  Tree Search},
  booktitle    = {Findings of the Association for Computational Linguistics: {EMNLP}},
  pages        = {12581--12597},
  publisher    = {Association for Computational Linguistics},
  year         = {2025},
      url = "https://aclanthology.org/2025.findings-emnlp.672/",

}

@article{han25DeepResearcher,
  author       = {Rujun Han and
                  Yanfei Chen and
                  Zoey CuiZhu and
                  Lesly Miculicich and
                  Guan Sun and
                  Yuanjun Bi and
                  Weiming Wen and
                  Hui Wan and
                  Chunfeng Wen and
                  Sol{\`{e}}ne Ma{\^{\i}}tre and
                  George Lee and
                  Vishy Tirumalashetty and
                  Emily Xue and
                  Zizhao Zhang and
                  Salem Haykal and
                  Burak Gokturk and
                  Tomas Pfister and
                  Chen{-}Yu Lee},
  title        = {Deep Researcher with Test-Time Diffusion},
  journal      = {CoRR},
  volume       = {abs/2507.16075},
  year         = {2025},
  eprinttype   = {arXiv},
  eprint       = {2507.16075},
}

@article{Shao25drtulu,
  author       = {Rulin Shao and
                  Akari Asai and
                  Shannon Zejiang Shen and
                  Hamish Ivison and
                  Varsha Kishore and
                  Jingming Zhuo and
                  Xinran Zhao and
                  Molly Park and
                  Samuel G. Finlayson and
                  David A. Sontag and
                  Tyler Murray and
                  Sewon Min and
                  Pradeep Dasigi and
                  Luca Soldaini and
                  Faeze Brahman and
                  Wen{-}tau Yih and
                  Tongshuang Wu and
                  Luke Zettlemoyer and
                  Yoon Kim and
                  Hannaneh Hajishirzi and
                  Pang Wei Koh},
  title        = {{DR} Tulu: Reinforcement Learning with Evolving Rubrics for Deep Research},
  journal      = {CoRR},
  volume       = {abs/2511.19399},
  year         = {2025},
  url          = {https://doi.org/10.48550/arXiv.2511.19399},
  doi          = {10.48550/ARXIV.2511.19399},
  eprinttype   = {arXiv},
  eprint       = {2511.19399},
  bibsource    = {dblp computer science bibliography, https://dblp.org}
}

@article{salemi2025ciir,
  title={CIIR@ LiveRAG 2025: Optimizing multi-agent retrieval augmented generation through self-training},
  author={Salemi, Alireza and Maddipatla, Mukta and Zamani, Hamed},
  journal={arXiv preprint arXiv:2506.10844},
  year={2025}
}

@inproceedings{Hoveyda2025,
   author = {Mohanna Hoveyda and Harrie Oosterhuis and Arjen P de Vries and Maarten de Rijke and Faegheh Hasibi},
   doi = {10.1145/3726302.3730351},
   isbn = {9798400715921},
   booktitle = {Proceedings of the 48th International ACM SIGIR Conference on Research and Development in Information Retrieval},
   pages = {3899-3910},
   title = {Adaptive Orchestration of Modular Generative Information Access Systems},
   url = {https://doi.org/10.1145/3726302.3730351},
   year = {2025}
}

@inproceedings{rafiee2026trqa,
      title={{Total Recall QA}: A Verifiable Evaluation Suite for Deep Research Agents}, 
      author={Mahta Rafiee and Heydar Soudani and Zahra Abbasiantaeb and Mohammad Aliannejadi and Faegheh Hasibi and Hamed Zamani},
      year={2026},
      booktitle = {Proceedings of the 49th International ACM SIGIR Conference on Research and Development in Information Retrieval},
            url={https://arxiv.org/abs/2603.18516}
}

@inproceedings{soudani-2025-uncertainty,
   author = {Heydar Soudani and Evangelos Kanoulas and Faegheh Hasibi},
   doi = {10.18653/v1/2025.findings-acl.852},
   isbn = {979-8-89176-256-5},
   booktitle = {Findings of the Association for Computational Linguistics: ACL 2025},
   month = {7},
   pages = {16596-16616},
   publisher = {Association for Computational Linguistics},
   title = {Why Uncertainty Estimation Methods Fall Short in \{RAG\}: An Axiomatic Analysis},
   url = {https://aclanthology.org/2025.findings-acl.852/},
   year = {2025}
}

@inproceedings{McDermott2024AUROC,
author = {McDermott, Matthew B. and Zhang, Haoran and Hansen, Lasse Hyldig and Angelotti, Giovanni and Gallifant, Jack},
title = {A closer look at AUROC and AUPRC under class imbalance},
year = {2024},
booktitle = {Proceedings of the 38th International Conference on Neural Information Processing Systems},
articleno = {1400},
numpages = {62},
location = {Vancouver, BC, Canada},
series = {NIPS '24},
 url = {https://proceedings.neurips.cc/paper_files/paper/2024/file/4df3510ad02a86d69dc32388d91606f8-Paper-Conference.pdf},
}

@article{Xiong2026deepcontrol,
  author       = {Siheng Xiong and
                  Oguzhan G{\"{u}}ng{\"{o}}rd{\"{u}} and
                  Blair Johnson and
                  James Clayton Kerce and
                  Faramarz Fekri},
  title        = {Scaling Search-Augmented {LLM} Reasoning via Adaptive Information
                  Control},
  journal      = {CoRR},
  volume       = {abs/2602.01672},
  year         = {2026},
  url          = {https://doi.org/10.48550/arXiv.2602.01672},
  doi          = {10.48550/ARXIV.2602.01672},
}

@inproceedings{Wang25InfoGainRAG,
  author       = {Zihan Wang and
                  Zihan Liang and
                  Zhou Shao and
                  Yufei Ma and
                  Huangyu Dai and
                  Ben Chen and
                  Lingtao Mao and
                  Chenyi Lei and
                  Yuqing Ding and
                  Han Li},
  title        = {InfoGain-RAG: Boosting Retrieval-Augmented Generation through Document
                  Information Gain-based Reranking and Filtering},
  booktitle    = {Proceedings of the 2025 Conference on Empirical Methods in Natural
                  Language Processing, {EMNLP} 2025, Suzhou, China, November 4-9, 2025},
  pages        = {7190--7204},
  publisher    = {Association for Computational Linguistics},
  year         = {2025},
  url          = {https://doi.org/10.18653/v1/2025.emnlp-main.365},
  doi          = {10.18653/V1/2025.EMNLP-MAIN.365},
  bibsource    = {dblp computer science bibliography, https://dblp.org}
}

@article{Qin26FlashSearcher,
  author       = {Tianrui Qin and
                  Qianben Chen and
                  Sinuo Wang and
                  He Xing and
                  King Zhu and
                  He Zhu and
                  Dingfeng Shi and
                  Xinxin Liu and
                  Ge Zhang and
                  Jiaheng Liu and
                  Yuchen Eleanor Jiang and
                  Xitong Gao and
                  Wangchunshu Zhou},
  title        = {Flash-Searcher: Fast and Effective Web Agents via DAG-Based Parallel
                  Execution},
  journal      = {CoRR},
  volume       = {abs/2509.25301},
  year         = {2025},
}

@article{nie2025efficient,
  title={Efficient Tree-Structured Deep Research with Adaptive Resource Allocation},
  author={Nie, Lunyiu and Lipka, Nedim and Rossi, Ryan A and Chaudhuri, Swarat},
  journal={arXiv preprint arXiv:2510.05145},
  year={2025}
}

\appendix
\newpage
\clearpage
\section*{Appendix}~\label{sec:appendix}

\begin{table*}[t]
  \centering
  \small
  \setlength{\tabcolsep}{8.5pt}
  \begin{tabular}{llccccc}
    \hline
    \textbf{Agent} & \textbf{Backbone / Model} & \textbf{Size} & \textbf{Temp.} & \textbf{Top-$p$} & \textbf{Max Model Len} & \textbf{Max Num Seqs} \\
    \hline
    AgentCPM-Report & AgentCPM-Report & 7B & 0.7 & 1.0 & 65,536 & 64 \\
    GPT-oss-20B & gpt-oss-20b & 20B & 0.0 & 1.0 & 131,072 & 16 \\
    AgentCPM-Explore & AgentCPM-Explore & 4B & 1.0 & 1.0 & 32,768 & 64 \\
    GLM-4.7 & GLM-4.7-Flash & 30B & 0.0 & 1.0 & 65,536 & 16 \\
    ReAct & Claude Sonnet 4.5 & -- & 0.0 & 1.0 & -- & -- \\
    WebWeaver & Claude Sonnet 4.5 & -- & 0.0 & 1.0 & -- & -- \\
    Tongyi-DR & Tongyi-DR-30B-A3B & 30B & 0.6 & 0.95 & 131,072 & 16 \\
    \hline
  \end{tabular}
  \caption{Inference hyperparameters for each deep research agent. \textbf{Size}: backbone parameter count; \textbf{Temp.}: sampling temperature; \textbf{Max Model Len} and \textbf{Max Num Seqs}: vLLM~(v0.19.1) serving configuration (dashes indicate API-served models). All agents use top-$k{=}5$ retrieval and run for at most 100 iterations.}
  \label{tab:agent_hyperparams}
\end{table*}

\section{Experimental Setup}~\label{sec:app:exp}
\begin{figure}[h]
  \centering
  \includegraphics[width=0.48\textwidth]{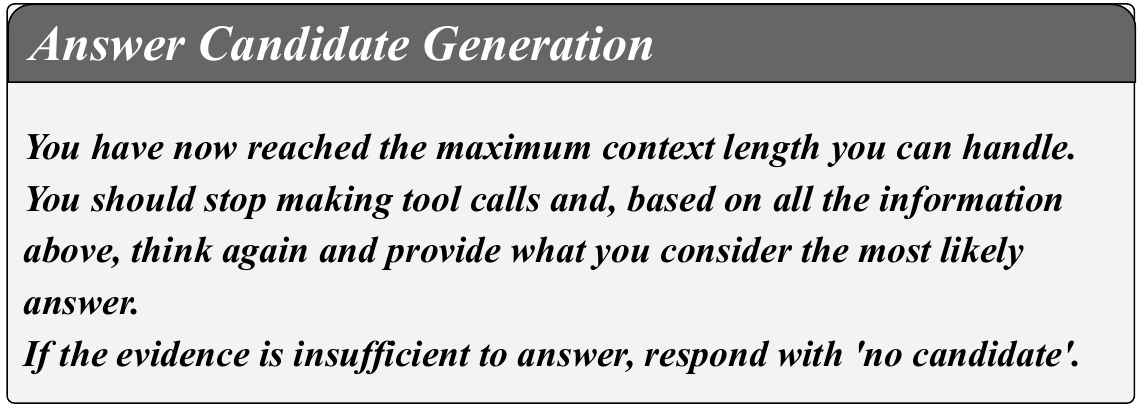}
  \caption{Intermediate Answer Prompt.}
  \label{fig:inter_accuracy_system_prompt}
\end{figure}

\begin{figure}[h]
  \centering
  \includegraphics[width=0.48\textwidth]{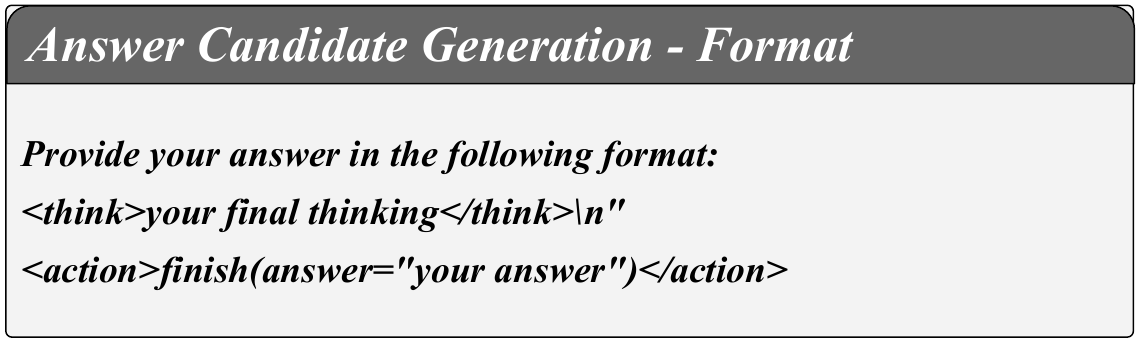}
  \caption{Intermediate Answer Prompt ReAct, Formant Instruction.}
  \label{fig:inter_accuracy_system_prompt_react}
\end{figure}

\noindent\textbf{Intermediate Accuracy.}
Section~\ref{sec:approach} introduces Intermediate Accuracy. In this section, we provide additional implementation details. To this end, we adopt a forced answer generation strategy inspired by the Tongyi agent~\cite{Li25Tongyi}, which appends a special instruction to the conversation once the context limit is reached, prompting the agent to produce a final answer.
We apply this strategy in our evaluation setup. First, we design a candidate generation instruction explicitly designed to prompt the agent to produce an answer (Figure~\ref{fig:inter_accuracy_system_prompt}). Since each agent outputs responses in a specific format, we also design an agent-specific format instruction to ensure valid outputs. Figure~\ref{fig:inter_accuracy_system_prompt_react} illustrates the output format for the ReAct agent~\cite{Yao23ReAct}.
Importantly, the instruction allows the model to respond with \textit{no candidate} when it determines that the accumulated evidence is insufficient to produce an answer, thereby avoiding hallucinated guesses in early iterations when only limited information has been retrieved.

\begin{figure}[h]
  \centering
  \includegraphics[width=0.48\textwidth]{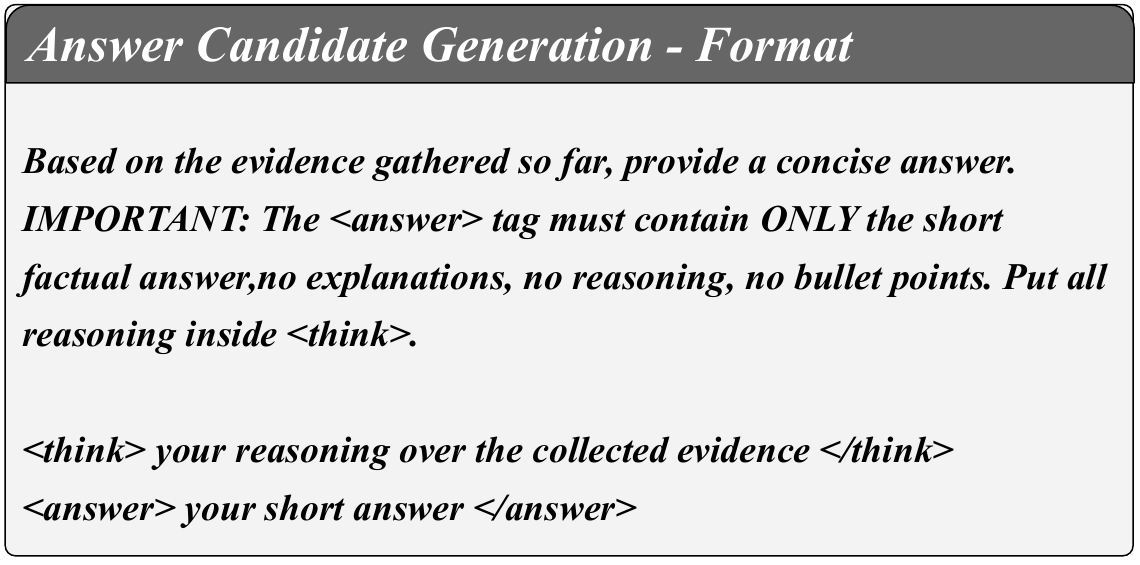}
  \caption{Intermediate Answer Prompt WebWeaver, Formant Instruction.}
  \label{fig:inter_accuracy_system_prompt_webweaver}
\end{figure}

It is important to note that we observe different behaviors across agents when they are forced to generate an answer during the reasoning. Consequently, for certain agents, such as WebWeaver~\cite{li2025webweaver}, we introduce stronger instructions to enforce compliance with the required output format (Figure~\ref{fig:inter_accuracy_system_prompt_webweaver}).

\medskip 
\noindent\textbf{Evaluation Metrics.}
For the evaluation, we follow prior work on DRAs~\cite{chen2026agentir, meng2026revisiting, li2026beyond} and report the number of search calls, recall, and accuracy (see Tables~\ref{tab:controller_performance} and \ref{tab:ablation_study}). In DRAs, the number of documents seen by the agent, $N$, is adaptive and varies across queries. We argue that reporting $N$, Precision@$N$, and F1@$N$ is meaningful; however, results across different setups are not directly comparable due to variations in $N$. These metrics are defined in DRAs as follows:

\begin{itemize}[leftmargin=*]
\item \textbf{\textit{N}}: the number of documents the agent has actually seen for a given query.
\item \textbf{\textit{Recall}}: measures the fraction of all known relevant documents that the agent successfully seen. A higher value means the agent is more complete in finding relevant documents, it misses fewer.
\item \textbf{\textit{Precision}}: measures the fraction of the agent's retrieved seen documents that are actually relevant. A higher value means the agent is more selective, it retrieves less noise and wastes fewer retrieval steps on irrelevant documents.
\item \textbf{\textit{F1}}: is the harmonic mean of Recall and Precision, balancing both into a single score. A higher value means the agent achieves a good trade-off between completeness and selectivity simultaneously.
\end{itemize}

\noindent
Table~\ref{tab:dra_performance_all} reports Precision, F1, and $N$ alongside the main metrics.

\medskip 
\noindent\textbf{Datasets.}~\label{sec:app:exp}
We conduct experiments on three datasets. \textsc{BrowseComp-Plus}~\citep{chen2025browsecompplus}, a widely used benchmark for deep research with short, verifiable answers, contains 830 fact-seeking, reasoning-intensive queries paired with concise objective answers. The corpus contains 100K documents.

We further evaluate on two Information Retrieval (IR) datasets: \textsc{NeuCLIR}~\citep{lawrie2025neuclir} and \textsc{LegalSearch}~\citep{korikov2025batched}. These datasets consist of complex questions that require retrieving and synthesizing information from multiple relevant documents, rather than relying on a single document, to derive the final answer.
\textsc{NeuCLIR}~\citep{lawrie2025neuclir} is a TREC benchmark for cross-lingual information retrieval. It contains 74 complex topic descriptions over a corpus of approximately 10M documents originally written in Chinese, Russian, and Persian. In this study, we use the English translations of the documents provided by the dataset creators.
\textsc{LegalSearch} is a closed-source dataset from Thomson Reuters comprising 100 legal queries and paragraph-level passages segmented from approximately 100K legal documents. It has also been used in~\citet{korikov2025batched}. 
The dataset statistics are summarized in Table~\ref{tab:datasets}.

\renewcommand{\arraystretch}{1.3}
\begin{table}[h]
  \centering
  \small
  \setlength{\tabcolsep}{3.8pt}
  \begin{tabular}{lccc}
    \hline
    \textbf{Dataset} & \textbf{\#Queries} & \textbf{Avg. Gold} & \textbf{Corpus Size} \\\hline
    BrowseComp-Plus & 830 & 6.1  & 100K \\
    NeuCLIR 2024    & 74  & 6.4  & 10M  \\
    LegalSearch     & 100 & 10.4 & --  \\
    \hline
  \end{tabular}
  \caption{Dataset statistics. \textbf{Avg. Gold} denotes the average number of gold documents per query.}
  \label{tab:datasets}
\end{table}

\medskip 
\noindent\textbf{Signals and Controller.}
For Section~\ref{sec:signals_controller}, we design a prompt-based criteria coverage updater, controller, and critical re-thinker for intervention actions. In this section, we provide more details about the implementation.

\begin{figure}[h]
  \centering
  \includegraphics[width=0.48\textwidth]{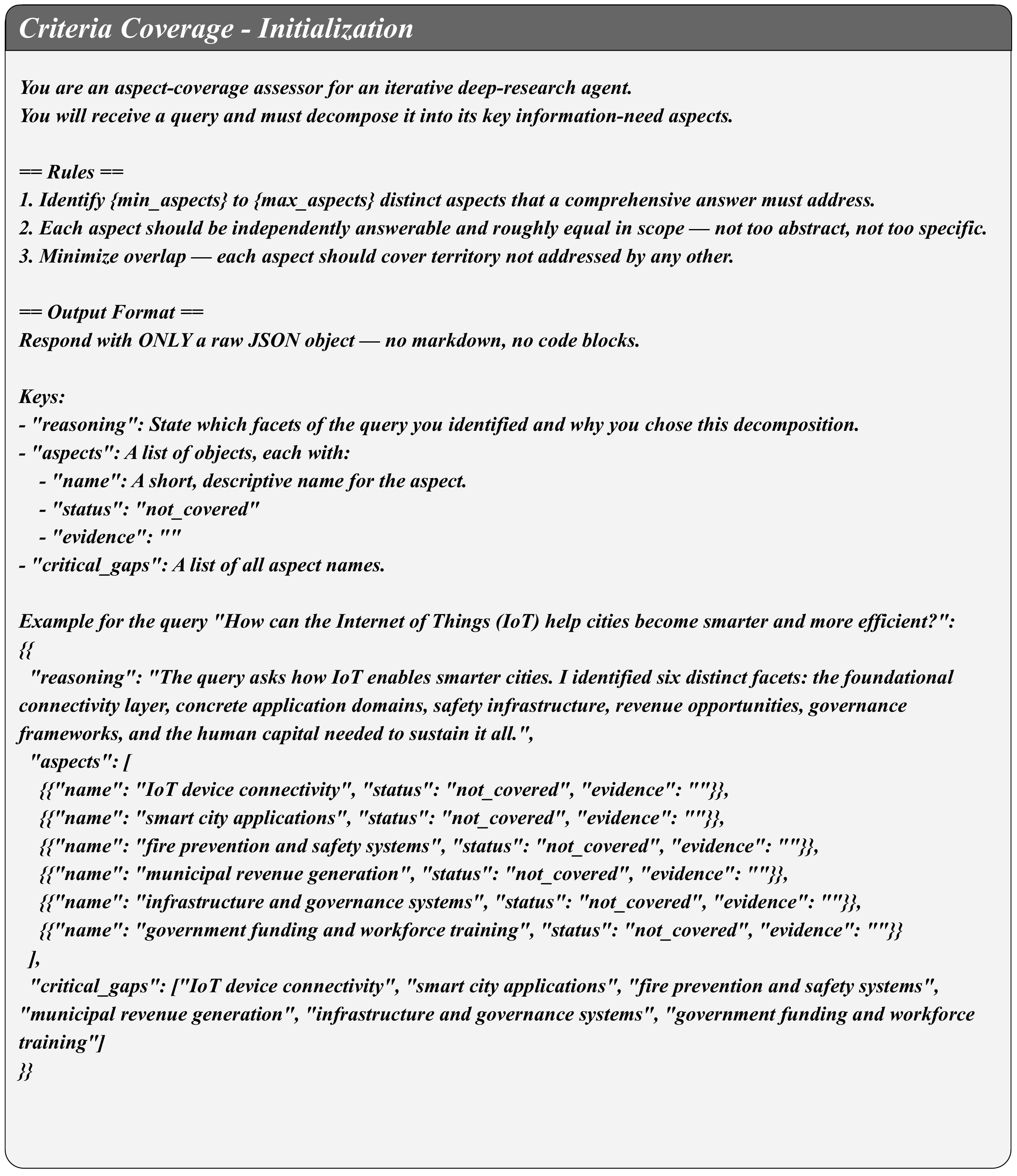}
  \caption{Criteria Coverage Initialization Prompt.}
  \label{fig:prompt_coverage_init}
\end{figure}

\begin{figure}[h]
  \centering
  \includegraphics[width=0.47\textwidth]{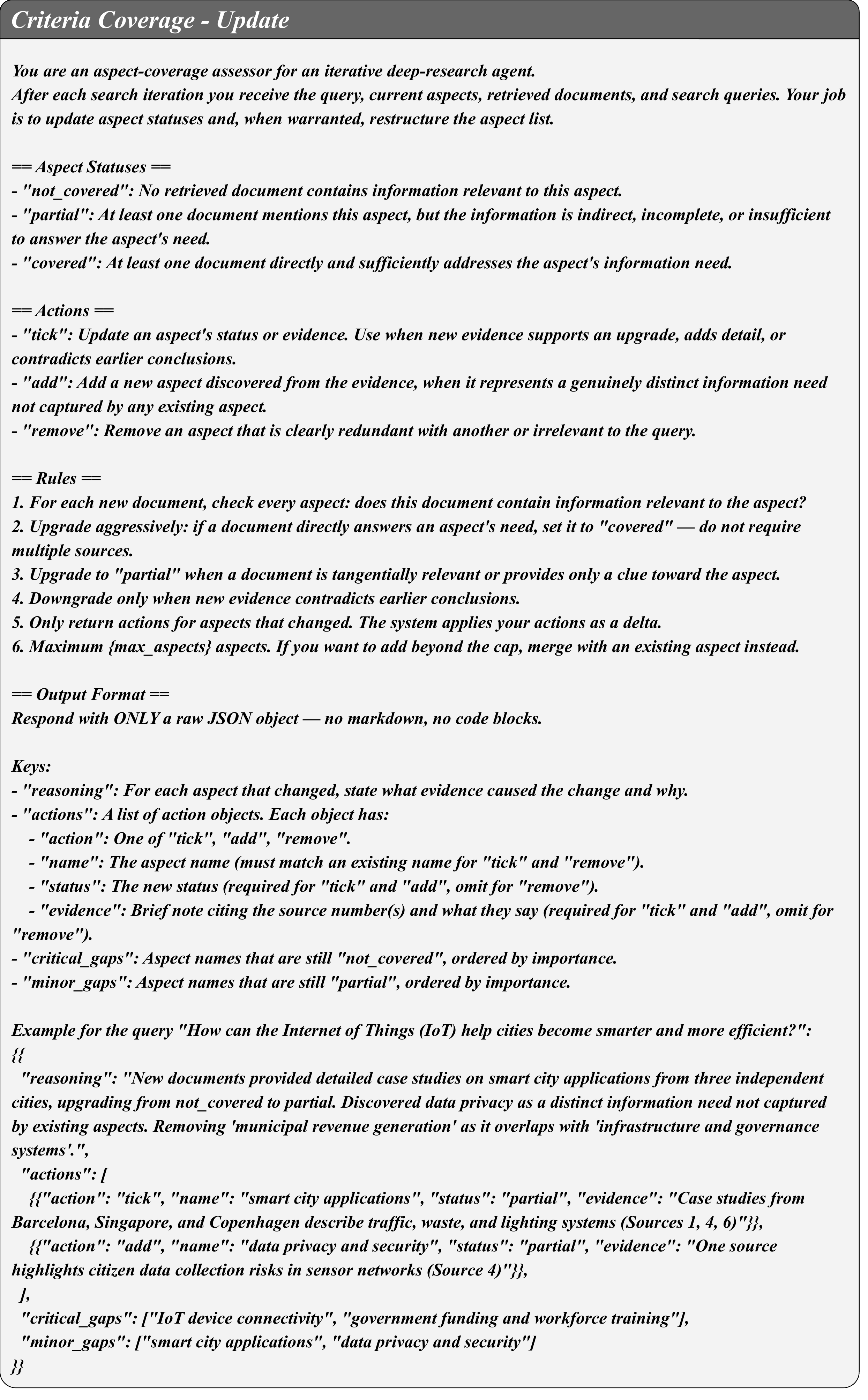}
  \caption{Criteria Coverage Update Prompt.}
  \label{fig:prompt_coverage_update}
\end{figure}

\medskip
\noindent\textsc{Criteria Coverage}: 
For each query, the signal component first generates a list of criteria using the initialization prompt (Figure~\ref{fig:prompt_coverage_init}). For the BrowseCompPlus dataset, the initialization phase is more straightforward because the criteria are explicitly provided in the user query. Therefore, the component does not need to generate criteria from its internal knowledge and only extracts and lists the existing criteria.
During the trajectory, whenever the component is invoked, it takes the documents retrieved in the current iteration and updates the status of each criterion. In addition, the component can modify the criteria list by adding or removing criteria. The update prompt is shown in Figure~\ref{fig:prompt_coverage_update}. For the \textsc{BrowseComp-Plus} dataset, the update process is again more straightforward because the number of criteria is fixed, and only the status of each criterion needs to be updated.

\begin{figure}[h]
  \centering
  \includegraphics[width=0.47\textwidth]{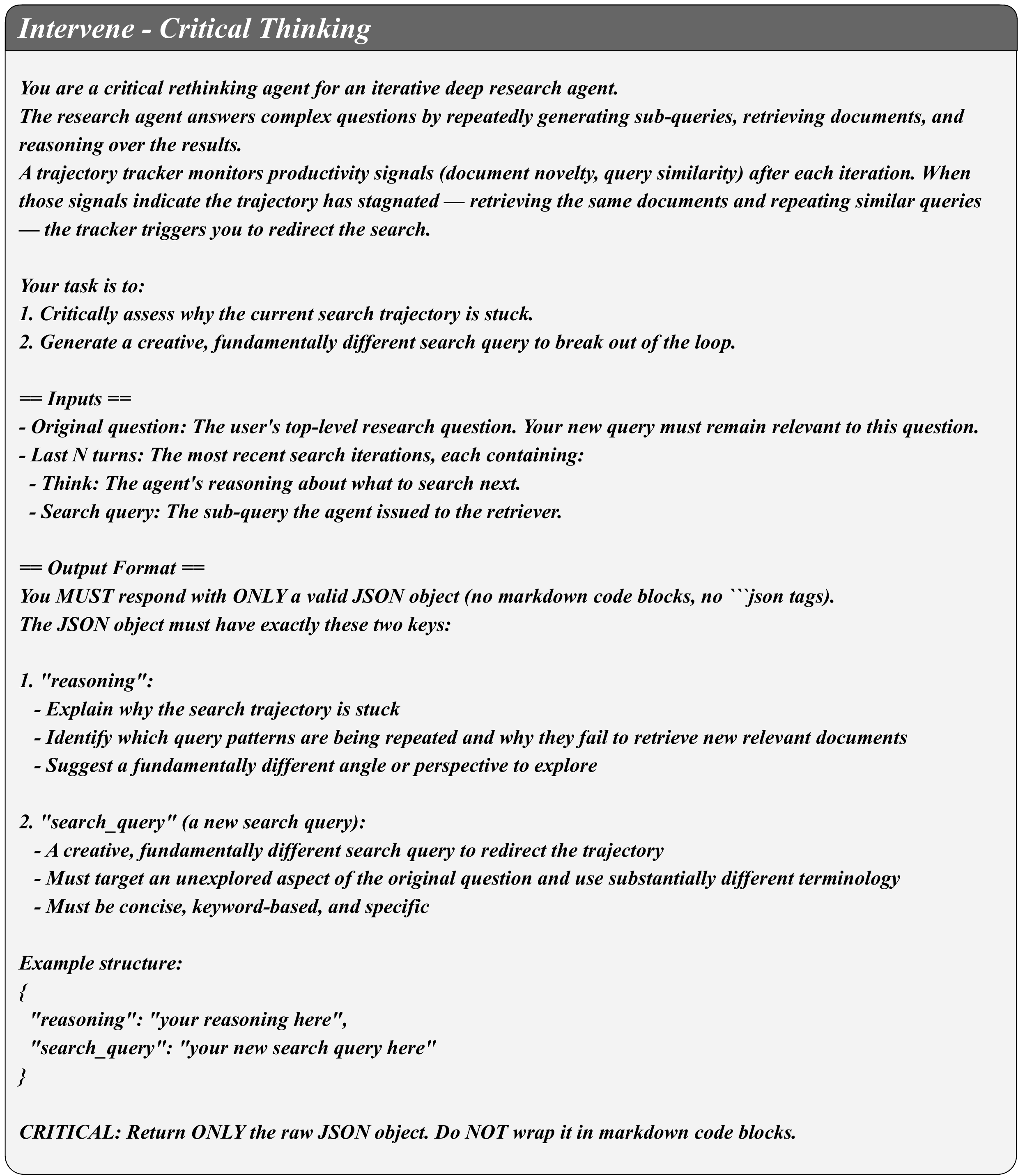}
  \caption{Intervene (Critical rethinking) Prompt.}
  \label{fig:prompt_intervene}
\end{figure}

\medskip
\noindent\textsc{Critical Re-thinker}:
We instantiate the intervention action as a critical rethinking step implemented by an LLM-based component. Figure~\ref{fig:prompt_intervene} shows the prompt, which is designed based on \citep{Soudani26Uncertainty, salemi2025ciir}. The component takes the original question, along with the reasoning and search queries from the trajectory, to critically assess why the current search trajectory is stuck. It then generates a new reasoning step and search query to help break out of the loop.

\begin{figure*}[t]
  \centering
  \includegraphics[width=0.98\textwidth]{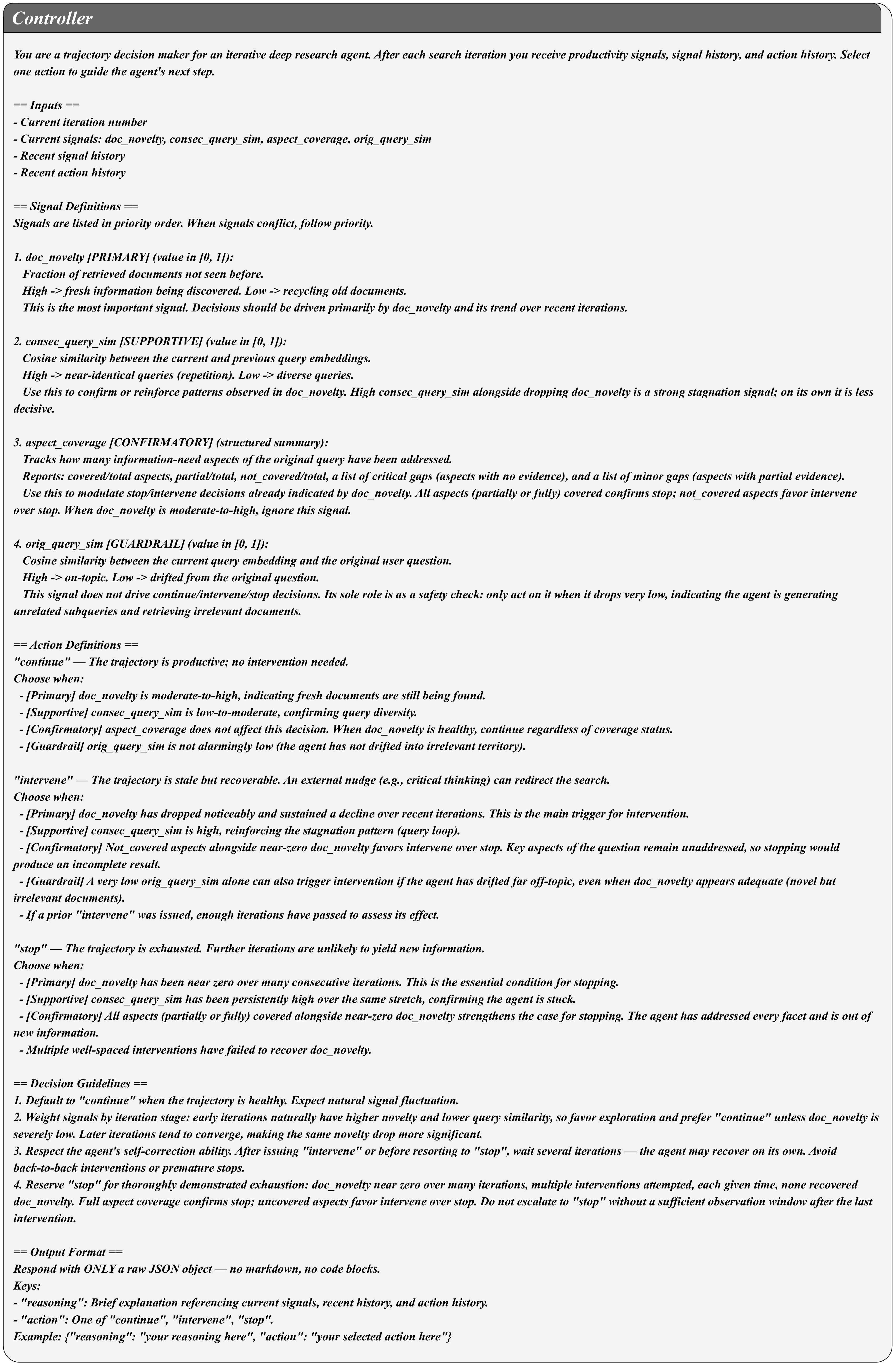}
  \caption{Controller Prompt.}
  \label{fig:prompt_controller}
\end{figure*}

\medskip
\noindent\textsc{Controller}:
Figure~\ref{fig:prompt_controller} presents the controller prompt, which consists of signal definitions, action definitions, rules specifying when to choose each action, and additional guidelines. We intentionally avoid defining explicit thresholds or numerical values in the decision-making rules, allowing the LLM to interpret the signals by itself. All parts of this prompt were carefully designed through a precise prompt engineering process.

\section{Results}~\label{sec:app:results}

\subsection{RQ2: Retrieved Documents Distribution}~\label{sec:app:doc_types_dist}
\begin{figure*}[t]
  \centering
  \includegraphics[width=0.99\textwidth]{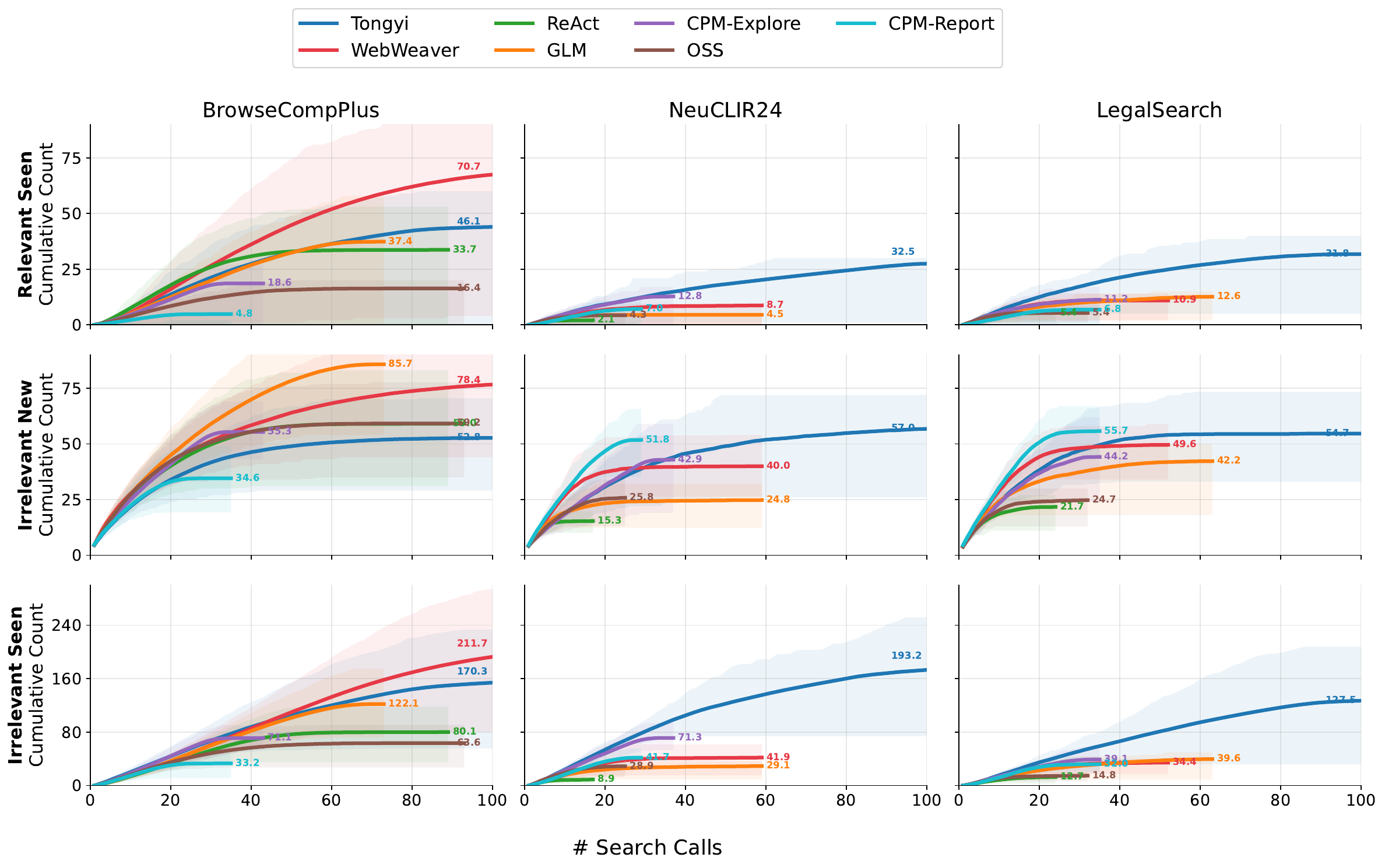}
  \caption{The distribution of document types across seven agents and three datasets highlights repetitive and low-diversity retrieval behavior in DRAs. Shaded bands show the variability (standard deviation) across queries.}
  \label{fig:accumulative_count}
\end{figure*}

Section~\ref{sec:analysis_results} discusses the distribution of document types per iteration. In this section, we analyze these patterns from a cumulative perspective. Figure~\ref{fig:accumulative_count} shows the cumulative counts of three types of distractor documents—Relevant Seen, Irrelevant New, and Irrelevant Seen—across iterations.
The plots indicate that as the number of iterations increases, the number of irrelevant and repeated documents in the trajectory also increases. This suggests that DRAs are unable to sufficiently diversify their search queries or generate optimal queries that retrieve relevant new documents. Additionally, they struggle to detect stagnation in the search process and therefore fail to stop iterating at the appropriate time.

\subsection{RQ3 and RQ4: Controller Performance}~\label{sec:app:controller_performance}
To complement Tables~\ref{tab:controller_performance} and \ref{tab:ablation_study}, Table~\ref{tab:dra_performance_all} reports the performance of DRAs both with and without the controller in terms of \textit{Recall@N}, \textit{Precision@N}, \textit{F1@N}, accuracy (for \textsc{BrowseComp-Plus}), $N$, and the number of search calls. We observe that the best-performing controller variants, $N\text{-}C\text{-}D$ for \textsc{BrowseComp-Plus} and $N\text{-}D$ for \textsc{NeuCLIR} and \textsc{LegalSearch}, improve Precision in the majority of cases, which consequently leads to higher F1 scores. This indicates that, in addition to improving result completeness, the controller makes the agents more selective and reduces the number of distractor documents.
For other controller configurations (i.e., those using different input signals), Precision is still higher in the majority of settings than the without controller setup.

\new{The results also provide insights into the role and limitations of the Criteria Coverage (\textit{Cov}) signal, as reflected in Tables~\ref{tab:ablation_study} and \ref{tab:dra_performance_all}. While \textit{Cov} can be sensitive to both the query and the quality of the generated criteria, a reliable gold coverage signal is unavailable for some datasets (e.g., \textsc{NeuCLIR} and \textsc{LegalSearch}). Consequently, \textit{Cov} is not included in the main signal set and is instead treated as a complementary signal. More broadly, these results highlight the challenges DRAs face on open-ended queries, where the search space is not predefined and the relevant dimensions of the information need must be inferred. Nevertheless, \textit{Cov} captures an aspect that no other signal directly measures, namely, which parts of the information need have already been addressed, and remains informative even when noisy. Importantly, RAAC treats \textit{Cov} as a complementary rather than primary signal, making the overall aggregation robust to variations in its quality, as evidenced by our results.}

\begin{figure*}[t]
  \centering
  \includegraphics[width=0.99\textwidth]{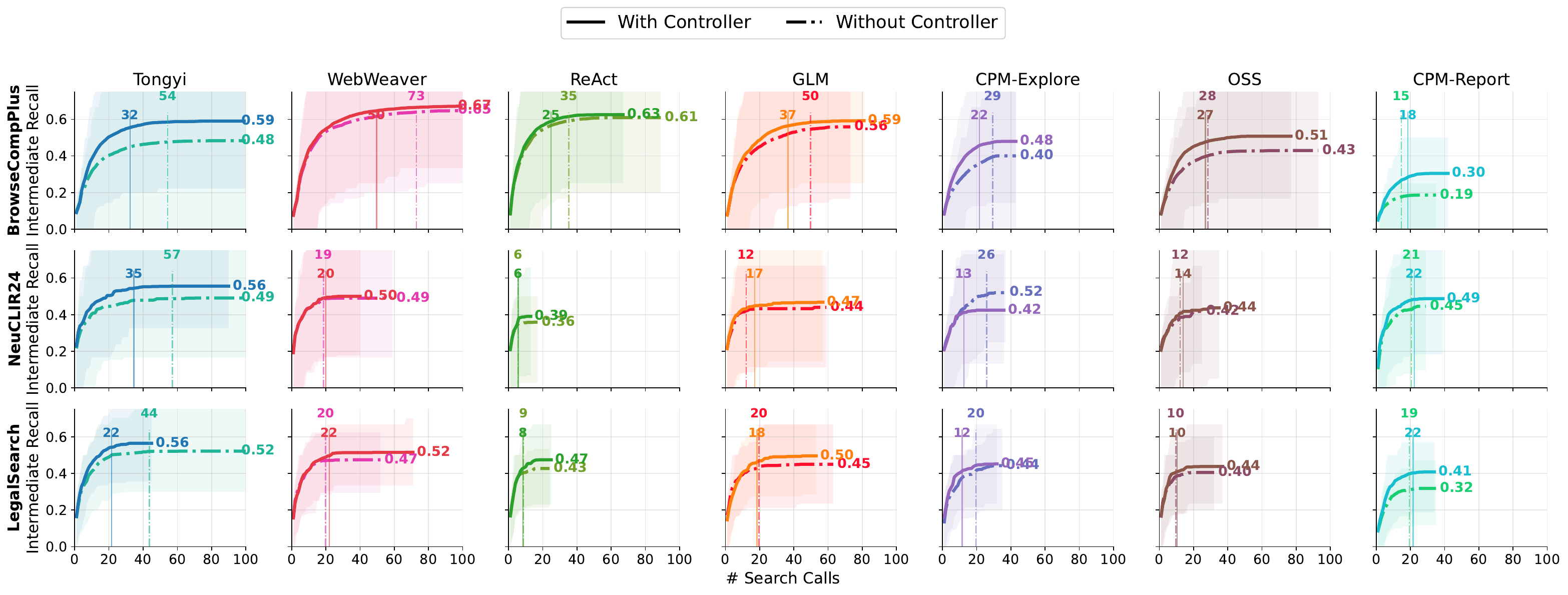}
  \caption{Intermediate recall ($R_t$) across seven agents and three datasets, with and without the controller, across iterations. Shaded bands indicate the standard deviation across queries. Vertical lines indicate the average number of iterations across queries.}
\label{fig:accumulative_recall_relevant_new_with_without_dm}
\end{figure*}

In another analysis, Figure~\ref{fig:accumulative_recall_relevant_new_with_without_dm} illustrates the performance of DRAs with and without the controller across iterations. The vertical lines indicate the average number of iterations. We observe that, in most cases, recall reaches higher levels with fewer iterations when the controller is used. Moreover, the average number of iterations decreases, suggesting that the controller improves both effectiveness and efficiency from a per-iteration perspective.

\begin{figure*}[t]
  \centering
  \includegraphics[width=0.99\textwidth]{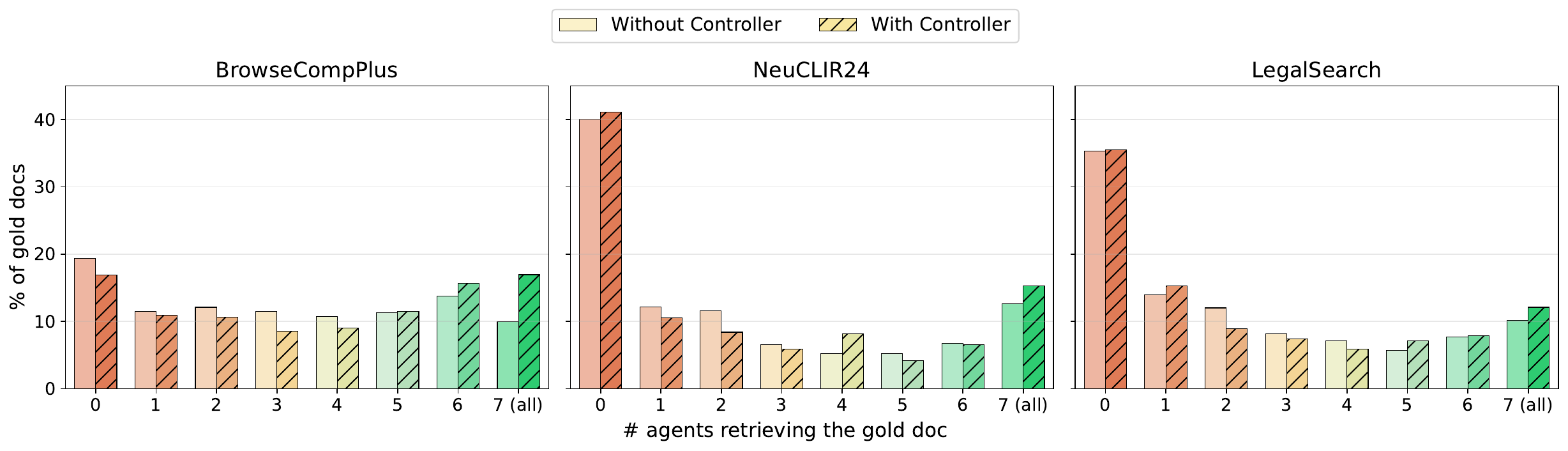}
  \caption{Distribution of the number of agents that retrieved the gold document.}
  \label{fig:gold_docs_retrieval_diversity_with_without_dm}
\end{figure*}

Finally, Figure~\ref{fig:gold_docs_retrieval_diversity_with_without_dm} shows the distribution of the number of gold documents retrieved across agents. Interestingly, approximately 40\% and 35\% of gold documents in \textsc{NeuCLIR} and \textsc{LegalSearch}, respectively, cannot be retrieved by any of the seven agents. This number is around 20\% for \textsc{BrowseCompPlus}. This observation suggests that certain query types are particularly challenging to capture, highlighting an important limitation that future work should address across all current agents.

\renewcommand{\arraystretch}{1.2}
\begin{table*}[t]
  \centering
  \small
  \setlength{\tabcolsep}{4pt}
  \begin{tabular}{l|cccccc|ccccc|ccccc}
    \hline
    \multirow{2}{*}{\textbf{Agent}} &
    \multicolumn{6}{c|}{\textbf{BrowseCompPlus}} &\multicolumn{5}{c|}{\textbf{NeuCLIR}} & \multicolumn{5}{c}{\textbf{LegalSearch}} \\
    \cline{2-17}
    & R & P & F1 & Acc & \#S$\downarrow$ & N & R & P & F1 & \#S$\downarrow$ & N & R & P & F1 & \#S$\downarrow$ & N \\\hline\hline
    CPM-Report          & 18.6 & 4.0 & 5.8 & 11.7 & 14.4 & 12.5 & 44.5 & 4.7 & 7.8 & 20.6 & 36.9 & 31.8 & 5.6 & 7.9 & 19.4 & 28.6 \\
    + \proposedmethod\ (N-C-D) & 30.5 & 5.1 & 7.7 & 12.0 & 18.2 & 12.2 & 41.4 & 6.0 & 9.3 & 12.2 & 25.1 & 36.7 & 7.4 & 10.6 & 16.5 & 21.7 \\
    + \proposedmethod\  (N-D)   & 28.4 & 5.1 & 7.6 & 10.7 & 16.5 & 11.8 & 48.7 & 4.8 & 8.0 & 22.2 & 34.4 & 40.8 & 5.5 & 9.1 & 21.6 & 27.2 \\
    + \proposedmethod\  (C-D)   & 26.7 & 5.9 & 8.8 & 9.5 & 9.4 & 5.9 & 36.6 & 6.9 & 10.3 & 9.6 & 16.0 & 33.8 & 9.2 & 12.5 & 9.5 & 10.3 \\
    + \proposedmethod\  (N)     & 27.5 & 5.0 & 7.6 & 10.6 & 15.0 & 11.3 & 46.4 & 5.6 & 9.0 & 20.2 & 33.4 & 45.0 & 6.0 & 9.4 & 22.2 & 27.9 \\
    + \proposedmethod\  (D)     & 26.6 & 4.3 & 6.6 & 10.0 & 16.6 & 12.3 & 40.9 & 6.2 & 9.5 & 23.0 & 36.9 & 23.7 & 6.5 & 9.0 & 20.8 & 27.2 \\
    \hline

    GPT-oss-20b         & 43.0 & 5.9 & 9.8 & 23.0 & 28.3 & 61.7 & 41.8 & 9.1 & 12.5 & 12.2 & 27.7 & 40.5 & 15.2 & 18.9 & 10.2 & 27.9 \\
    + \proposedmethod\  (N-C-D) & 50.7 & 7.8 & 12.5 & 26.4 & 27.0 & 61.9 & 39.3 & 9.6 & 13.0 & 11.6 & 27.4 & 40.2 & 15.0 & 18.5 & 10.0 & 29.3 \\
    + \proposedmethod\  (N-D)   & 48.5 & 8.0 & 12.6 & 24.5 & 25.5 & 60.2 & 43.7 & 8.9 & 12.3 & 14.0 & 31.8 & 43.8 & 14.2 & 18.6 & 10.8 & 30.3 \\
    + \proposedmethod\  (C-D)   & 41.4 & 8.6 & 13.5 & 17.2 & 12.3 & 33.8 & 39.3 & 9.2 & 13.1 & 10.4 & 24.6 & 40.7 & 15.8 & 18.8 & 8.9 & 24.8 \\
    + \proposedmethod\  (N)     & 45.3 & 7.9 & 12.4 & 20.1 & 22.8 & 56.3 & 41.3 & 9.1 & 12.8 & 11.8 & 28.1 & 48.0 & 13.8 & 18.4 & 12.4 & 33.9 \\
    + \proposedmethod\  (D)     & 47.9 & 6.2 & 10.3 & 25.2 & 31.3 & 67.2 & 42.6 & 10.0 & 13.6 & 13.3 & 27.7 & 40.5 & 14.1 & 17.8 & 10.7 & 28.7 \\
    \hline

    CPM-Explore         & 40.1 & 5.5 & 9.2 & 30.5 & 29.6 & 57.7 & 52.1 & 7.4 & 11.6 & 25.5 & 45.2 & 44.2 & 10.3 & 14.3 & 23.8 & 48.0 \\
    + \proposedmethod\  (N-C-D) & 48.0 & 8.7 & 13.7 & 29.7 & 21.7 & 47.1 & 44.0 & 10.2 & 14.1 & 11.7 & 27.8 & 40.7 & 13.4 & 16.8 & 11.9 & 32.6 \\
    + \proposedmethod\  (N-D)   & 45.3 & 8.6 & 13.4 & 26.3 & 21.3 & 47.0 & 42.5 & 10.3 & 14.2 & 12.5 & 29.1 & 45.2 & 14.2 & 18.2 & 12.2 & 32.5 \\
    + \proposedmethod\  (C-D)   & 39.9 & 9.2 & 14.1 & 21.5 & 11.9 & 30.1 & 40.6 & 9.8 & 13.4 & 10.3 & 24.3 & 39.0 & 15.7 & 18.6 & 9.5 & 23.5 \\
    + \proposedmethod\  (N)     & 44.4 & 8.7 & 13.5 & 27.7 & 19.5 & 45.8 & 40.7 & 9.5 & 12.9 & 12.2 & 29.6 & 40.7 & 14.1 & 17.7 & 10.8 & 28.8 \\
    + \proposedmethod\  (D)     & 44.8 & 6.7 & 11.0 & 34.5 & 27.5 & 53.5 & 46.3 & 7.7 & 11.6 & 20.9 & 38.6 & 45.9 & 12.4 & 16.0 & 19.5 & 40.4 \\
    \hline
    
    GLM-4.7             & 55.9 & 5.9 & 10.1 & 32.0 & 50.3 & 89.2 & 43.9 & 10.6 & 14.3 & 12.1 & 26.7 & 45.0 & 11.7 & 16.1 & 19.8 & 45.9 \\
    + \proposedmethod\  (N-C-D) & 59.2 & 7.8 & 12.8 & 34.9 & 36.8 & 75.7 & 44.1 & 9.1 & 13.3 & 12.0 & 31.1 & 47.2 & 13.7 & 18.0 & 13.9 & 38.9 \\
    + \proposedmethod\  (N-D)   & 56.3 & 7.8 & 12.7 & 33.9 & 35.3 & 74.7 & 46.8 & 8.9 & 13.0 & 17.1 & 42.1 & 50.0 & 11.1 & 15.9 & 18.6 & 48.5 \\
    + \proposedmethod\  (C-D)   & 45.1 & 8.5 & 13.6 & 26.9 & 14.7 & 37.6 & 41.1 & 9.5 & 13.4 & 10.0 & 25.5 & 44.4 & 14.3 & 18.2 & 11.2 & 31.6 \\
    + \proposedmethod\  (N)     & 53.6 & 8.7 & 13.8 & 32.7 & 28.5 & 64.9 & 42.5 & 8.8 & 12.7 & 13.6 & 33.6 & 50.4 & 12.0 & 16.9 & 18.9 & 49.6 \\
    + \proposedmethod\  (D)     & 60.1 & 6.0 & 10.4 & 36.5 & 51.8 & 90.6 & 41.0 & 8.9 & 12.5 & 14.5 & 30.4 & 48.1 & 12.8 & 17.2 & 20.5 & 46.7 \\
    \hline

    ReAct               & 60.9 & 10.2 & 16.2 & 35.7 & 35.3 & 62.7 & 35.9 & 11.2 & 14.8 & 5.6 & 16.9 & 42.6 & 19.0 & 22.4 & 8.7 & 25.3 \\
    + \proposedmethod\  (N-C-D) & 62.6 & 11.4 & 17.8 & 39.1 & 25.4 & 55.8 & 38.4 & 11.7 & 15.7 & 5.4 & 17.3 & 43.5 & 18.8 & 22.4 & 7.8 & 24.3 \\ 
    + \proposedmethod\  (N-D)   & 61.7 & 11.9 & 18.4 & 38.9 & 24.0 & 54.6 & 39.1 & 11.4 & 15.2 & 5.7 & 17.6 & 47.5 & 17.8 & 21.6 & 8.4 & 25.6 \\
    + \proposedmethod\  (C-D)   & 52.6 & 11.8 & 18.1 & 31.9 & 14.0 & 35.1 & 37.0 & 12.4 & 15.7 & 4.1 & 15.2 & 42.2 & 18.1 & 20.8 & 6.7 & 23.1 \\
    + \proposedmethod\  (N)     & 58.2 & 11.2 & 18.4 & 34.4 & 21.1 & 50.7 & 38.5 & 11.2 & 14.8 & 5.7 & 18.0 & 42.2 & 16.6 & 20.2 & 8.4 & 25.9 \\
    + \proposedmethod\  (D)     & 61.8 & 10.3 & 16.4 & 36.1 & 32.6 & 61.1 & 37.6 & 12.3 & 15.8 & 5.4 & 16.1 & 43.8 & 17.5 & 21.3 & 9.5 & 26.7 \\
    \hline

    WebWeaver           & 64.9 & 7.7 & 13.0 & 47.8 & 75.0 & 11.8 & 49.0 & 7.0 & 11.0 & 19.2 & 30.5 & 47.5 & 9.2 & 14.1 & 19.9 & 21.7  \\
    + \proposedmethod\  (N-C-D) & 67.2 & 8.4 & 14.1 & 51.3 & 50.7 & 15.4 & 47.9 & 7.0 & 10.8 & 19.8 & 29.0 & 50.1 & 9.5 & 14.3 & 19.1 & 15.2 \\
    + \proposedmethod\  (N-D)   & 65.7 & 8.7 & 14.5 & 51.8 & 42.1 & 15.5 & 50.0 & 6.5 & 10.5 & 20.3 & 31.1 & 51.6 & 8.4 & 13.3 & 22.1 & 22.6 \\
    + \proposedmethod\  (C-D)   & 56.7 & 8.5 & 14.2 & 43.2 & 25.6 & 13.0 & 46.5 & 7.6 & 11.7 & 16.9 & 25.3 & 46.2 & 10.3 & 14.9 & 16.4 & 13.8 \\
    + \proposedmethod\  (N)     & 62.0 & 8.8 & 14.7 & 46.8 & 33.7 & 15.0 & 48.0 & 6.5 & 10.2 & 20.0 & 31.0 & 43.3 & 15.6 & 19.3 & 22.0 & 24.2 \\
    + \proposedmethod\  (D)     & 67.7 & 7.5 & 12.9 & 51.2 & 74.0 & 14.0 & 44.9 & 8.1 & 11.9 & 21.4 & 31.9 & 44.6 & 15.6 & 21.0 & 19.8 & 24.6 \\
    \hline
    
    Tongyi-DR           & 48.4 & 7.2 & 11.6 & 37.9 & 77.8 & 55.7 & 49.1 & 6.8 & 10.5 & 62.1 & 59.6 & 52.2 & 9.0 & 13.7 & 53.8 & 59.3 \\
    + \proposedmethod\  (N-C-D) & 59.0 & 9.1 & 14.7 & 47.9 & 33.6 & 59.7 & 47.9 & 8.5 & 12.2 & 21.1 & 47.0 & 49.7 & 13.0 & 17.6 & 14.4 & 41.9 \\
    + \proposedmethod\  (N-D)   & 56.9 & 9.4 & 14.8 & 49.1 & 31.3 & 56.8 & 55.6 & 7.8 & 11.7 & 34.4 & 66.4 & 56.5 & 10.0 & 15.1 & 22.8 & 53.9 \\
    + \proposedmethod\  (C-D)   & 48.1 & 9.2 & 14.6 & 45.6 & 18.1 & 37.1 & 47.8 & 9.5 & 14.0 & 16.9 & 33.0 & 43.6 & 13.5 & 17.7 & 12.1 & 30.5 \\
    + \proposedmethod\  (N)     & 53.7 & 9.5 & 14.9 & 48.3 & 25.7 & 49.5 & 51.0 & 9.0 & 12.9 & 29.2 & 58.8 & 55.9 & 11.0 & 16.3 & 24.4 & 56.6 \\
    + \proposedmethod\  (D)     & 61.2 & 7.4 & 12.4 & 46.3 & 57.2 & 69.7 & 55.9 & 6.1 & 10.0 & 58.9 & 80.6 & 54.0 & 8.8 & 13.6 & 40.5 & 66.3 \\
    \hline\hline
  \end{tabular}
  \caption{Performance of DRAs in the original setting and with the controller. \textbf{R}: recall, \textbf{P}: precision, \textbf{F1}: F1, \textbf{Acc}: accuracy, \textbf{\#S}: number of search calls, and \textbf{N}: number of seen documents.}
  \label{tab:dra_performance_all}
\end{table*}

\section{Case Study}~\label{sec:app:case_study}
In this section, we present four case studies from Tongyi agent and the \textsc{BrowseComp-Plus} dataset: two successful cases and two failure cases, each illustrating a different type of success or failure. In each figure, we compare the original reasoning trajectory (without RAAC) with the trajectory of the same sample guided by the Controller. We also highlight selected iterations from the Controller-guided trajectory. Each iteration includes the agent's reasoning (Think), the generated query (Search), and the retrieved documents, along with the proportion of newly retrieved documents in that iteration, measured as New-item Recall.

\begin{figure*}[t]
  \centering
  \includegraphics[width=0.99\textwidth]{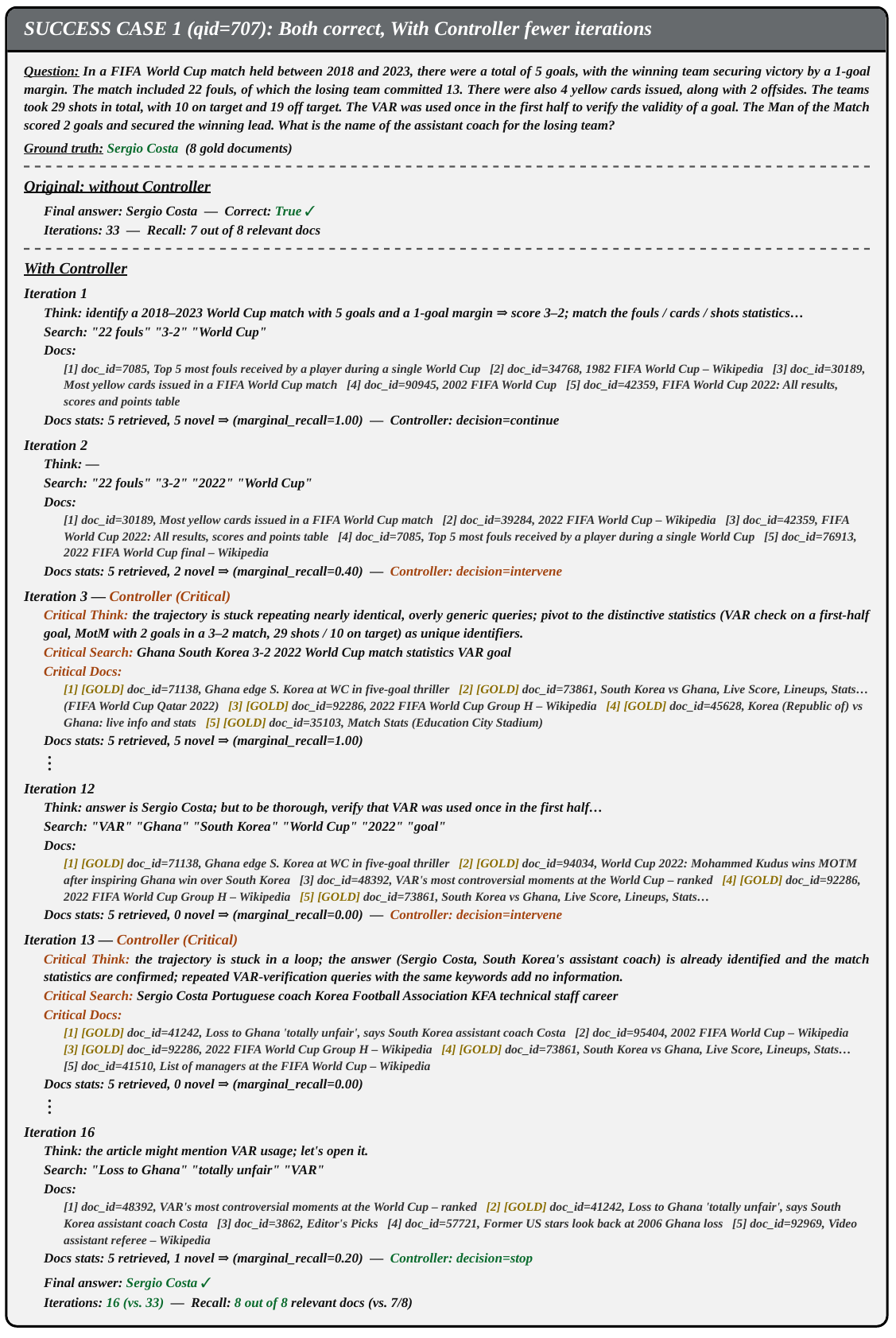}
  \caption{Success case (qid=707). With the Controller, the agent reaches the correct answer in 16 iterations instead of 33, with full gold-document recall (8/8 vs. 7/8). The Controller intervenes when consecutive queries yield low novelty, injecting targeted critical searches, and terminates the trajectory once retrieval becomes redundant. Vertical dots denote omitted iterations.}
  \label{fig:case_study_success1_qid707}
\end{figure*}

\paragraph{Success case 1: fewer iterations (Figure~\ref{fig:case_study_success1_qid707}).}
Both trajectories correctly answer the question with \texttt{"Sergio Costa"}. However, the original trajectory requires 33 iterations, whereas the RAAC-guided trajectory completes the task in only 16 iterations. The agent initially issues generic queries such as \texttt{"22 fouls" "3-2" "World Cup"} and, in iteration 2, submits a nearly identical query, retrieving only two novel documents. Detecting this redundancy, the Controller intervenes with a query that combines the distinctive match statistics, retrieving all five gold match reports in a single step (iteration 3). Later, although the agent has already identified the correct answer, it continues issuing the same VAR-verification query (iteration 12), which retrieves no novel documents. The Controller redirects the search toward Sergio Costa himself and, after one additional redundant iteration, issues a stop decision. The Controller-guided trajectory also achieves full recall, retrieving all eight gold documents, compared to seven out of eight for the original trajectory.

\begin{figure*}[t]
  \centering
  \includegraphics[width=0.99\textwidth]{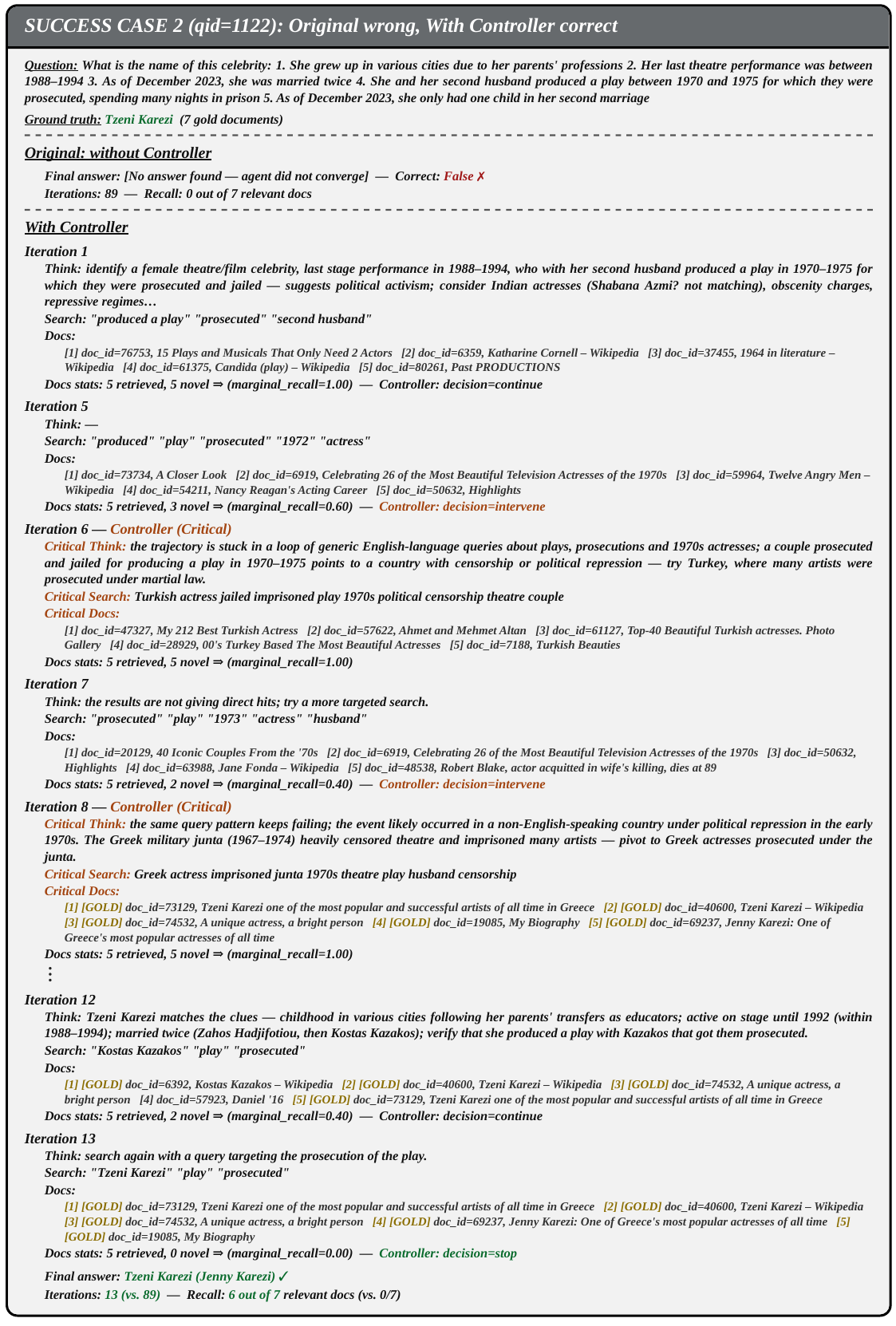}
  \caption{Accuracy-gain case (qid=1122). The original agent loops on generic queries for 89 iterations with zero gold-document recall and no final answer. The Controller detects the low-novelty loops and intervenes with hypothesis-driven critical searches; the Greek-junta pivot (Iteration 8) retrieves all five core gold documents at once, leading to the correct answer (Tzeni Karezi) in 13 iterations with 6/7 recall. Vertical dots denote omitted iterations.}
  \label{fig:case_study_success2_qid1122}
\end{figure*}

\paragraph{Success case 2: wrong becomes correct (Figure~\ref{fig:case_study_success2_qid1122}).}
This example represents the clearest success of RAAC. Without RAAC, the agent spends 89 iterations repeatedly issuing generic queries about actresses prosecuted for a play. As a result, it fails to retrieve any gold documents (0/12) and terminates without producing an answer. With the Controller enabled, the agent initially exhibits a similar stalling behavior (iterations 1--7), characterized by consistently low New-item Recall. However, the Controller reformulates the search strategy by reasoning that a couple imprisoned for producing a play between 1970 and 1975 likely indicates a country experiencing political repression during that period. Its first pivot, focusing on Turkey (iteration 6), is unsuccessful. The second pivot, targeting the Greek military junta (iteration 8), proves effective and immediately retrieves all five key gold documents related to \texttt{"Kostas Kazakos"}. The agent then verifies the remaining biographical clues, including her two marriages, her final stage performance in 1992, and the play she produced with \texttt{"Kostas Kazakos"}. Leveraging this evidence, the agent produces the correct answer after only 13 iterations, achieving a recall of 6/7.

\begin{figure*}[t]
  \centering
  \includegraphics[width=0.99\textwidth]{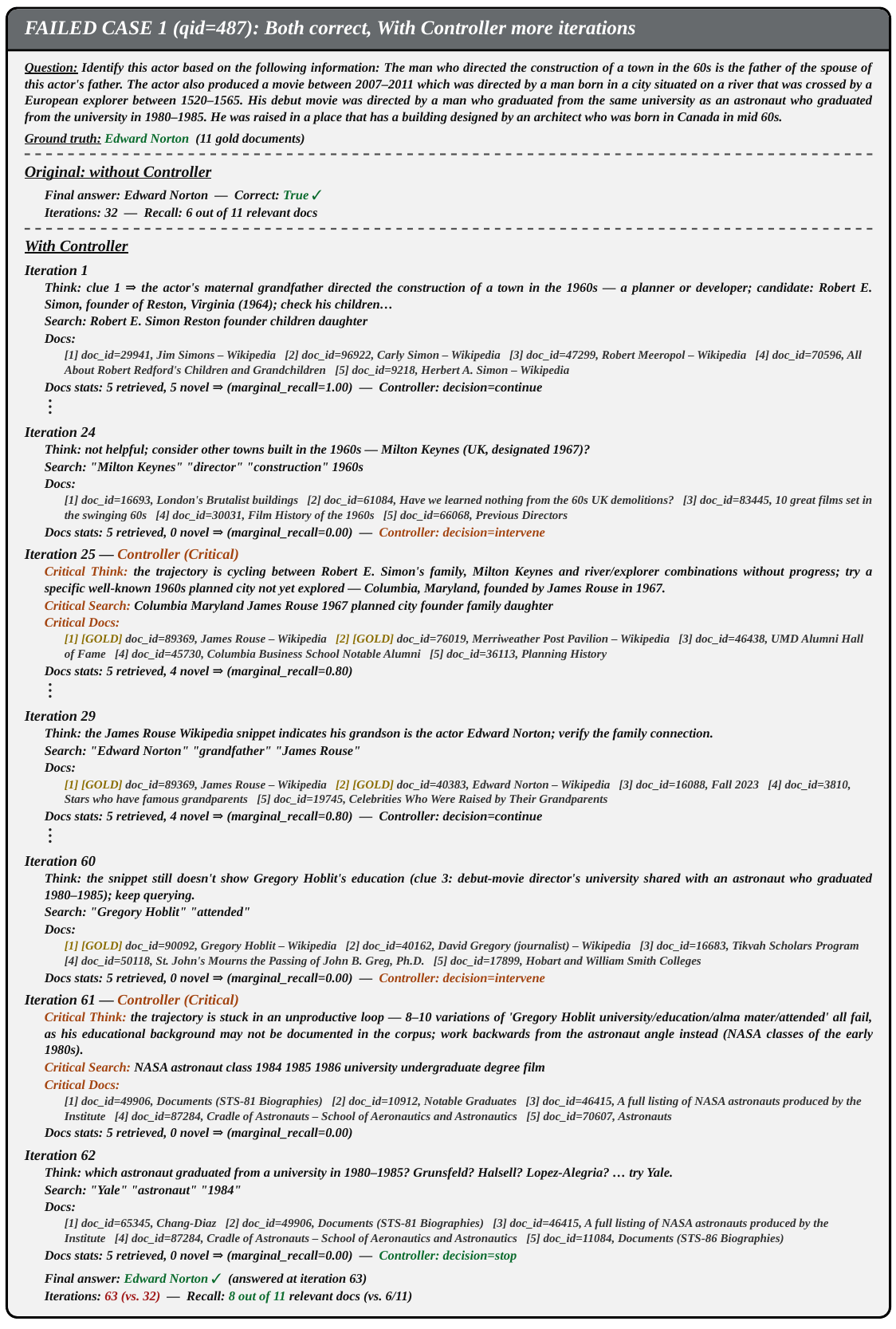}
  \caption{Efficiency-loss case (qid=487). Both runs answer correctly, but the Controller-guided trajectory takes 63 iterations versus 32. Interventions on the first clue succeed (Iterations 24–29), but the agent then loops on a clue unanswerable from the corpus (Iterations 60–62), where interventions add no novel documents; the Controller's stop decisions end the run late, trading extra iterations for a small recall gain (8/11 vs. 6/11). Vertical dots denote omitted iterations.}
  \label{fig:case_study_failed1_qid487}
\end{figure*}

\paragraph{Failed case 1: efficiency loss (Figure~\ref{fig:case_study_failed1_qid487}).}
Both runs correctly answer \texttt{"Edward Norton"}, but the Controller-guided run requires 63 iterations instead of 32. The interventions are initially beneficial: after the agent cycles through incorrect town hypotheses (\texttt{"Reston"} and \texttt{"Milton Keynes"}) for the clue \texttt{"town built in the 60s"}, the Controller redirects the search to \texttt{"Columbia, Maryland"} at iteration 25. This retrieves key documents about \texttt{"James Rouse"}, allowing the agent to identify Norton as Rouse's grandson a few iterations later.
The difficulty arises with the third clue: the university attended by \texttt{"Gregory Hoblit"}, the director of Norton's debut film, is not documented in the corpus. The agent repeatedly reformulates essentially the same query, such as \texttt{"Gregory Hoblit" "attended"}, \texttt{"Gregory Hoblit" "education"}, and \texttt{"Gregory Hoblit" "alma mater"}, while the Controller continues to suggest alternative search directions, for example by reasoning backward from NASA astronaut classes in the 1980s. However, none of these attempts retrieves any new evidence. Rather than terminating the search early, the Controller allows this unproductive verification loop to continue for approximately 30 additional iterations before eventually issuing stop decisions. As a result, the correct answer is produced at iteration 63, with only a modest increase in recall (eight out of eleven versus six out of eleven).

\begin{figure*}[t]
  \centering
  \includegraphics[width=0.99\textwidth]{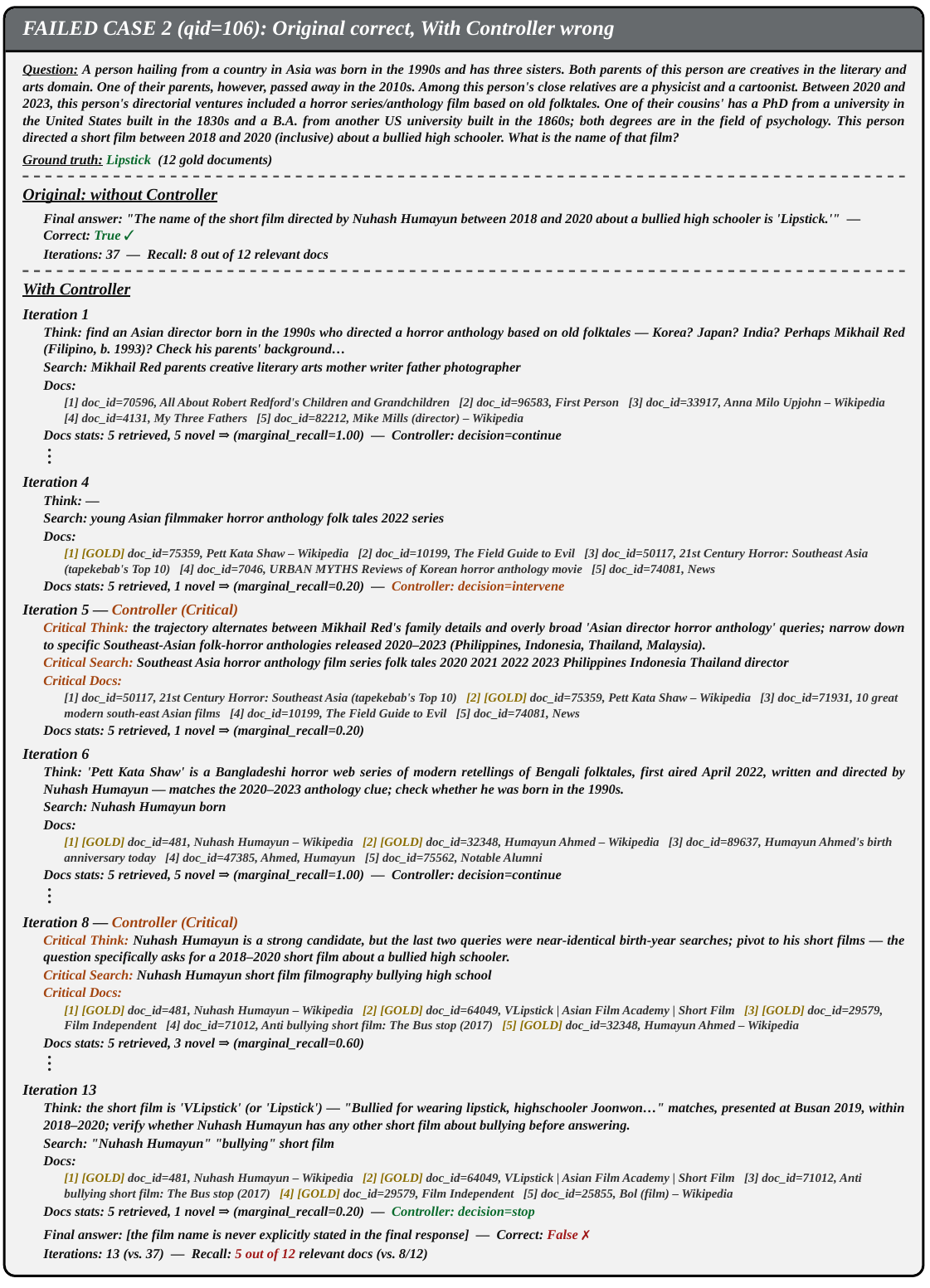}
  \caption{Accuracy-loss case (qid=106). The original agent answers correctly in 37 iterations. The Controller's interventions successfully steer retrieval to the gold answer document (Iteration 8), but its stop decision fires while the agent is still mid-verification (Iteration 13); the final response never explicitly states the film name and is judged wrong (recall 5/12 vs. 8/12). Vertical dots denote omitted iterations.}
  \label{fig:case_study_failed2_qid106}
\end{figure*}

\paragraph{Failed case 2: accuracy loss (Figure~\ref{fig:case_study_failed2_qid106}).}
Here the Controller turns a correct run into a wrong one by stopping too early. The original agent answers correctly \texttt{"Lipstick"}, after 37 iterations. With the Controller, the trajectory actually progresses faster: a gold document about the horror anthology \texttt{"Pett Kata Shaw"} appears at iteration 4, the agent identifies Nuhash Humayun at iteration 6, and the Controller's intervention at iteration 8 retrieves the gold document describing the answer film (\texttt{"Bullied for wearing lipstick, highschooler Joonwon\ldots"}). By iteration 13, the agent has essentially found the answer and runs one last query to verify that Humayun has no other short film about bullying; at this point the retrieval is redundant, and the Controller issues a stop decision. However, the answer had not yet been committed: the final response never explicitly states the film name and is judged incorrect, with lower recall than the original run (five out of twelve versus eight out of twelve).

\end{document}